\documentclass{egpubl}
\usepackage{subfiles}
\usepackage{soul}

\JournalSubmission    
\CGFccby

\usepackage[T1]{fontenc}
\usepackage{dfadobe}  

\biberVersion
\BibtexOrBiblatex
\usepackage[backend=biber,bibstyle=EG,citestyle=alphabetic,backref=true]{biblatex} 
\electronicVersion
\PrintedOrElectronic

\ifpdf \usepackage[pdftex]{graphicx} \pdfcompresslevel=9
\else \usepackage[dvips]{graphicx} \fi

\usepackage{egweblnk} 

\usepackage[nolist]{acronym}
\usepackage{xcolor}
\usepackage{amsmath}
\usepackage{booktabs}
\usepackage{multirow}
\usepackage{tikz}
\usepackage{xcolor}
\usepackage[export]{adjustbox}
\usepackage{lipsum}
\usepackage{placeins}
\usepackage{relsize}

\title[A Hierarchical Architecture for Neural Materials]%
      {A Hierarchical Architecture for Neural Materials}

\author[B. Xue \& S. Zhao\&H. W. Jensen\&Z. Montazeri]
{\parbox{\textwidth}{\centering B. Xue$^{1}$
        \qquad\qquad S. Zhao$^{2}$ \qquad\qquad H. W. Jensen$^{3}$ \qquad\qquad Z. Montazeri$^{1}$
        \\
        \footnotesize bowen.xue@postgrad.manchester.ac.uk \quad shz@ics.uci.edu \quad henrik.jensen@luxion.com \quad zahra.montazeri@manchester.ac.uk
        }
        \\
{\parbox{\textwidth}{\centering 
$^1$University of Manchester \quad$^2$University of California, Irvine\quad $^3$Luxion Inc}
}
}

\begin{document}

\teaser{
 \includegraphics[width=0.9\linewidth, trim=0 0 0 0, clip]{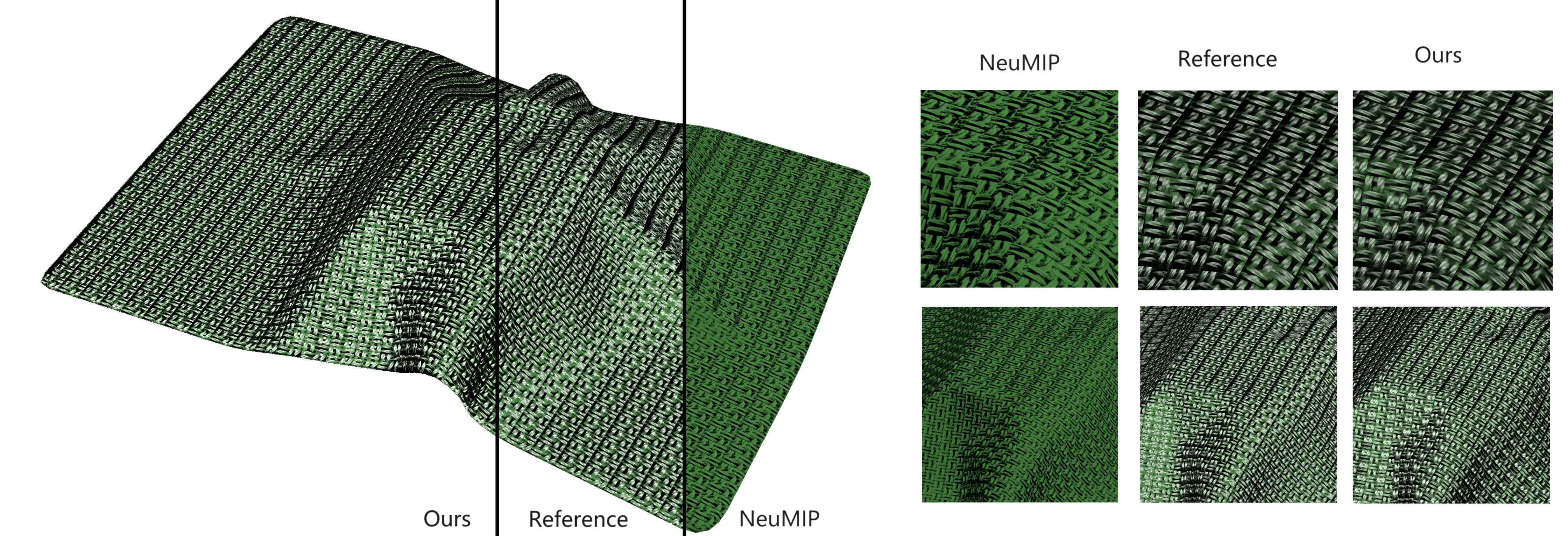}

 \centering
  \caption{Comparison of our method to the baseline NeuMIP model and the reference which is based on a 3D micro geometry of cloth. Our neural model enjoys 88\% lower MSE compared to NeuMIP in a relatively similar time and can accurately capture the self-shadowing as well as the sharp highlights of the complex materials which are known as the limitations of the original NeuMIP.}
\label{fig_teaser}
}
\maketitle
\begin{abstract}
Neural reflectance models are capable of reproducing the spatially-varying appearance of many real-world materials at different scales. Unfortunately, existing techniques such as NeuMIP have difficulties handling materials with strong shadowing effects or detailed specular highlights. In this paper, we introduce a neural appearance model that offers a new level of accuracy.
Central to our model is an inception-based core network structure that captures material appearances at multiple scales using parallel-operating kernels and ensures multi-stage features through specialized convolution layers. Furthermore, we encode the inputs into frequency space, introduce a gradient-based loss, and employ it adaptive to the progress of the learning phase. We demonstrate the effectiveness of our method using a variety of synthetic and real examples.




\end{abstract}  
\newcommand{\zm}[1]{\textcolor{blue}{Zahra:#1}}
\newcommand{\sz}[1]{\textcolor{green}{[SZ:#1]}}
\newcommand{\todo}[1]{\textcolor{red}{todo:#1}}

\newcommand*{\tabref}[1]{\tablename~\ref{#1}}
\newcommand*{\figref}[1]{Fig.~\ref{#1}}
\newcommand*{\tableref}[1]{Table~\ref{#1}}

\newcommand{\mypara}[1]{\noindent\textbf{#1:}\quad}

\newcommand{\mymath}[2]{
    \newcommand{#1}{\TextOrMath{$#2$\xspace}{#2}}}
\mymath{\decoder}{F}
\mymath{\image}{I}
\mymath{\loss}{\mathcal L}
\mymath{\kernel}{k}

\begin{acronym}
\acro{BRDF}{bidirectional reflection distribution function}
\acro{BTF}{bidirectional texture function}
\acro{MBTF}{mutli-scale BTF}
\acro{NeRF}{Neural Radiance Fields}
\end{acronym}

\section{Introduction}
\label{sec_intro}
Modeling the appearance of real-world materials in a physically faithful fashion is crucial for predictive rendering. This, however, is a challenging task: Many materials comprise complex fine-grained geometries that largely drive their macro-scale appearances.
Traditionally, material reflectance is specified using spatially varying BRDFs (SVBRDFs) or Bidirectional Texture Functions (BTFs). While these models work adequately for many applications, they are typically limited to one physical scale (or resolution). Further, SVBRDFs have difficulties handling parallax effects while BTFs \cite{dana1999} are highly data-intensive.

To address these limitations, several appearance models utilizing neural representations \cite{kuznetsov2021neumip, kuznetsov2022, fan2022neural} have been introduced recently. Although some of these methods---such as NeuMIP~\cite{kuznetsov2021neumip}---can be applied to learn the appearances of complex materials at varying scales, their physical accuracy can degrade drastically for material exhibiting complex shadows or specular highlights.


In this paper, we address this problem by introducing a neural appearance model with a new hierarchical architecture.
As shown in \figref{fig_teaser}, our model enjoys the generality to accurately capture complex directionally dependent effects---including shadows and highlights---at multiple physical scales.

Concretely, we make the following contributions:
\begin{itemize}
    \item We propose a new framework to improve neural materials to better capture highly glossy materials and better capture self-shadowing and sharp highlights by introducing a new hierarchical architecture and an input encoding step to the network to map the training inputs into a higher dimensional space~(\S\ref{ssec_network}).
    \item For better robustness, we also introduce new losses to allow our model to better capture both high- and low-frequency effects~(\S\ref{ssec_loss}).
\end{itemize}

We demonstrate the effectiveness of our technique by comparing it to the original NeuMIP~\cite{kuznetsov2021neumip} as shown in an example in \figref{fig_teaser}. In practice, similar to NeuMIP, our neural reflectance model can be integrated into most rasterization- and ray-tracing-based rendering systems.

\section{Related Work}
\label{sec_related}
%
 Neural rendering has emerged as a promising approach for a wide variety of applications, including material rendering, texture synthesis, and view synthesis. 
In this section, we review the most recent and relevant work in the area of neural rendering, focusing on techniques used for material rendering and displacement mapping.

\emph{Displacement mapping} serves as a powerful technique for augmenting material complexity on surface geometries, thereby yielding convincing parallax, silhouette, and shadowing effects. However, it imposes a considerable demand on computational resources. Conventional ray-tracing-based renderers typically effectuate displacement by tessellating the base geometry, a process that necessitates significant storage and computational capabilities \cite{thonat2021}.

A suite of techniques has been proposed as approximations to displacement mapping, including parallax mapping \cite{kaneko2001parallax}, relief mapping \cite{oliveira2000relief, policarpo2005relief}, view-dependent displacement mapping (VDM) \cite{wang2003vdm}, and generalized displacement maps (GDM) \cite{wang2004gdm, wang2005mdf}. These methodologies are fundamentally geometric and reliant on heightfields, overlooking the reflective effects emanating from the intricate material geometry.

\emph{Bidirectional Texture Function (BTF)} have been employed to represent arbitrary reflective surface appearances, first proposed by Dana et al. \cite{dana1999}. The storage of a discretized 6D function incurs substantial costs, and a multitude of compression techniques have been scrutinized \cite{filip2008compression}. Rainer et al. \cite{rainer2019autoencoder} introduced a neural architecture based on an autoencoder framework for compressing BTF slices per texel, and later advanced their work by integrating diverse materials into a shared latent space.

\emph{NeuMIP} \cite{kuznetsov2021neumip, kuznetsov2022}, an innovative neural approach, has been formulated to render and represent materials across disparate scales efficiently. Despite its advantages, NeuMIP faces constraints due to its network architecture and design, struggling to simulate the high-frequency information inherent in materials. Furthermore, it fails to accommodate curved surfaces. A more recent endeavor \cite{kuznetsov2022} aimed to overcome these shortcomings by incorporating surface curvature and transparency information into the neural model. Yet, the task of capturing high-frequency materials remains a formidable challenge. In this paper, we compare to NeuMIP \cite{kuznetsov2021neumip} and not the curved-surfaces variation \cite{kuznetsov2022} because our contribution is clearly visible in flat samples and our technique can be easily applicable to the newest version as well.

\emph{Recent advancements in \ac{NeRF}} have led to the development of methods capable of handling complex materials and geometries. The \ac{NeRF} technique \cite{mildenhall2020nerf}, which represents scenes as continuous volumetric fields, has been adapted for a plethora of applications, such as NeRF-W \cite{zhang2021nerfw} to manage view-dependent appearances, and Fourier plenOctrees for NeRF \cite{wang2022fourier} for real-time rendering. The methodology employed in this paper is inspired by \ac{NeRF} to capture details with sparse sampling. Rodriguez-Pardo et al. \cite{rodriguez2023neubtf} introduced a neural field-based framework for encoding and transferring Bidirectional Texture Functions (BTF). Baatz et al. \cite{baatz2022nerf} integrated Neural Reflectance Field Textures into the NeRF framework, enabling detailed rendering of complex materials. Lastly, Xiang et al. \cite{xiang2021neutex} demonstrated the use of neural networks for dynamic texture generation and mapping in volumetric neural rendering

\emph{Neural appearance modeling}
Fan et al. \cite{fan2022neural} introduced a universal decoder that can be applied to various materials and even to BRDFs not present in the training set. However, it requires a large neural network. Rainer et al. \cite{rainer2019autoencoder} proposed a neural structure based on an autoencoder framework to compress the BTF for each material. The decoder takes the latent vector and incoming and outgoing directions as inputs and each BTF requires separate training for the autoencoder. Later, Rainer et al. \cite{rainer2020unified} extended this work by suggesting a shared latent space for different materials. Xu et al. \cite{xu2023neusample} developed a novel importance sampling method for neural materials. Gauthier et al. \cite{gauthier2022mipnet} proposed a technique for mapping normal maps to anisotropic roughness levels. \cite{zeltner2023real} recently developed a neural rendering model using transformation layers and an encoder-decoder structure. However, due to the need for more parameters in the dataset, their model is not compatible with existing real BTF datasets. Furthermore, it does not support materials with displacement.

\emph{Micro-geometry appearance models} grapple with the granular details of the material and provide high-fidelity rendering results. The realistic rendering of fabrics, for instance, continues to be an elusive goal despite substantial efforts \cite{khungurn2015matching, Montazeri2021mechanics}. More recently, Montazeri et al. \cite{montazeri2020practical} introduced an efficient and unified shading model for woven and subsequently knit \cite{montazeri2021practical} fabrics, though these models do not address multi-resolution. In this study, we exploit their model to generate our fabric samples for training data.

\begin{figure}[t]
    \centering
    \hspace{-15pt}
    \includegraphics[width=0.5\textwidth]{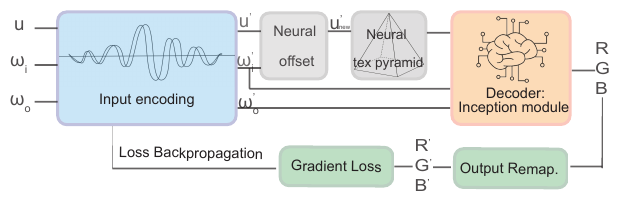}
    \caption{\textbf{Overview of our neural architecture.} The inputs are 2D spacial coordinates ($u$), incident and outgoing directions ($\omega_i$, and $\omega_o$), then encoded using Fourier transformation. The encoded $u'$ is again updated based on the micro-geometry using the Neural texture pyramid \cite{kuznetsov2021neumip}. We use the Inception modules illustrated in \figref{fig_architecture} to decode the color output. Lastly, we employ a remapping for optimized final output color ($R', G', B'$).
    \label{fig_overview}}
\end{figure}

\section{Our Method}
\label{sec_method}
In this section, we describe our neural method for modeling the appearances of complex materials exhibiting effects like shadows and specular highlights that cannot be accurately handled by previous neural models. 
Our model is inspired by NeuMIP \cite{kuznetsov2021neumip} and uses a novel network design.
Further, our technique is compatible with the more recent variant of NeuMIP \cite{kuznetsov2022} that utilizes displacement mapping for more accurate silhouettes.

In what follows, we first detail in \S\ref{ssec_network} our network design---which is crucial for better accuracy. Then, we explain in \S\ref{ssec_loss} our optimization strategies to further improve the accuracy of the training process. 

\subsection{Hierarchical Network Architecture}
\label{ssec_network}

\emph{Overview of NeuMIP.} 
The input to NeuMIP is a 7D parameter set including the position $u$, incoming and outgoing direction $\omega_i$, and $\omega_o$ as well as the kernel size for prefiltering. Their pipeline consisted of three main stages: (i) update the position $u$ to compensate the micro-geometry using a neural offset module; (ii) query a neural texture pyramid using the updated position to handle different levels of detail; and (iii) pass the queried feature vector to a decoder network to obtain the reflectance value. 

Our method uses the same three-step approach as NeuMIP but with several fundamental differences: we adopt the inception module decoder to replace the NeuMIP decoder, and the latent texture pyramid remains the same. This is shown in \figref{fig_overview} and detailed as follows.

\mypara{Inception module}
To better capture high-frequency effects such as detailed highlights or shading variations, we use an Inception module (instead of MLP layers used in NeuMIP). 

Inception modules \cite{inception} are specialized network blocks designed to approximate an optimal local structure of a convolutional network. It allows multiple types of filter sizes instead of a constant one. Networks leveraging the Inception modules \cite{inception} have been demonstrated capable of accurately preserving image features at both micro and macro scales.

\begin{figure}[t]
    \centering
    \hspace{-15pt}
    \includegraphics[width=0.5\textwidth]{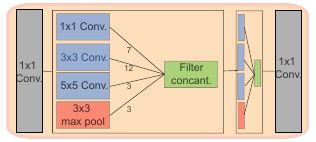}
    \caption{Our decoder architecture incorporates four layers of CNN, unlike NeuMIP that uses MLPs. The main distinction lies in the central incorporation of two Inception modules, flanked by two 1*1 convolution layers on both sides. This design choice significantly bolsters our performance due to its hierarchical structure.}
    \label{fig_architecture}
\end{figure}

Our core network architecture, which is predominantly based on the Inception modules, is demonstrated in \figref{fig_architecture}. The two $1 \times 1$ convolution layers at both ends serve the purpose of fully-connected layers and adjust the input and output size. Central to this design are the 4-layer Inception modules that capture the image features at four different scales using four kernel sizes that operate in parallel.For the convolutions in the inception module, they take all 25 channels as input.  Each of these Inception module blocks consists of four parallel pathways. The first three pathways employ convolutional layers with window sizes of $1 \times 1$, $3 \times 3$, and $5 \times 5$, respectively, to extract information at various spatial scales. The middle two pathways initially apply a $1 \times 1$ convolution to the input, which reduces the number of input channels and decreases the model complexity. The fourth pathway makes use of a $3 \times 3$ max-pooling layer, succeeded by a $1 \times 1$ convolutional layer to modify the number of channels. All four pathways introduce suitable padding to ensure that both input and output dimensions, in terms of height and width, remain consistent.

The output channel count for the Inception modules stands at $7 + 12 + 3 + 3 = 25$, with the output channel ratios for the four pathways being represented as $7:12:3:3 = 2:4:1:1$. Every Inception module is structured to support both an input and output of 25 channels. In our comparative study between fully connected networks of equivalent depth and neuron count, the fully connected networks displayed a noticeably lower performance. Moreover, ramping up the number of neurons or opting for deeper fully connected networks did not lead to notable enhancements in their ability to capture intricate details.

\begin{figure*}[ht!]
    \setlength{\tabcolsep}{1pt}
    \hspace{-18pt}
    \begin{tabular}{cccccc}
        w/o Inception
        & w/o Gradient loss  
        & w/o Input encoding
        & w/o Remapping  
        & Full model
        & Reference
        \\       
        \begin{tikzpicture}
            \node[anchor=south west,inner sep=0] (image) {\includegraphics[width=0.16\textwidth]{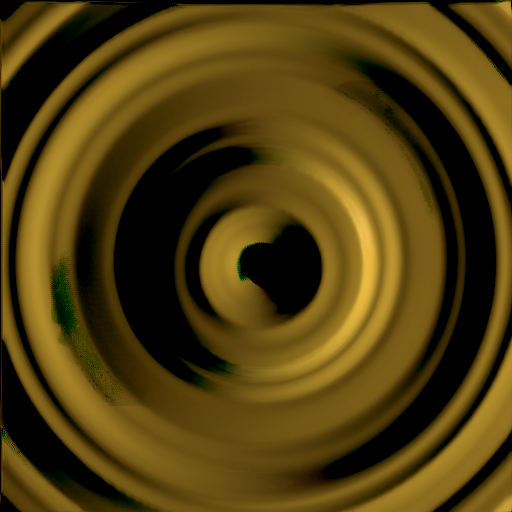}};
            \def\distanceTop{1.5cm};
            \def\distanceBottom{0.5cm};
            \def\distanceLeft{0.5cm};
            \def\distanceRight{1.5cm};
            \draw[red, thick] ([shift={(\distanceLeft,\distanceBottom)}]image.south west) rectangle ([shift={(-\distanceRight,-\distanceTop)}]image.north east);
        \end{tikzpicture}
        &
        \begin{tikzpicture}
            \node[anchor=south west,inner sep=0] (image) {\includegraphics[width=0.16\textwidth]{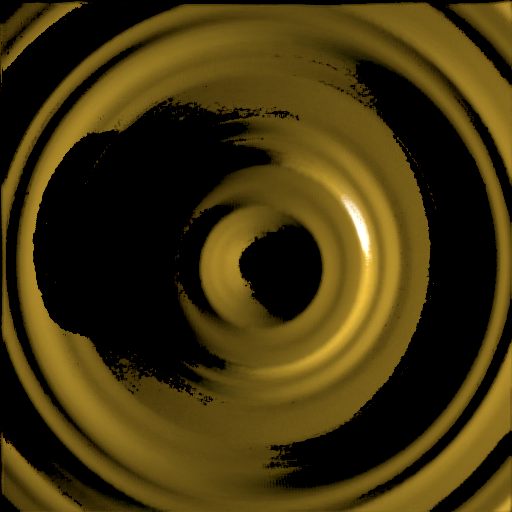}};
            \def\distanceTop{1.5cm};
            \def\distanceBottom{0.5cm};
            \def\distanceLeft{0.5cm};
            \def\distanceRight{1.5cm};
            \draw[red, thick] ([shift={(\distanceLeft,\distanceBottom)}]image.south west) rectangle ([shift={(-\distanceRight,-\distanceTop)}]image.north east);            
        \end{tikzpicture}
        &
        \begin{tikzpicture}
            \node[anchor=south west,inner sep=0] (image) {\includegraphics[width=0.16\textwidth]{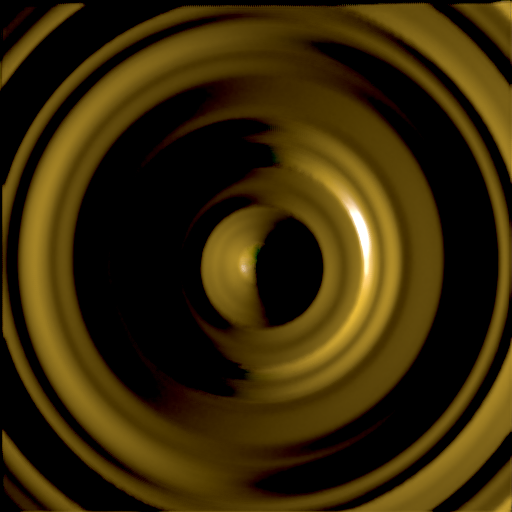}};
            \def\distanceTop{1.5cm};
            \def\distanceBottom{0.5cm};
            \def\distanceLeft{0.5cm};
            \def\distanceRight{1.5cm};
            \draw[red, thick] ([shift={(\distanceLeft,\distanceBottom)}]image.south west) rectangle ([shift={(-\distanceRight,-\distanceTop)}]image.north east);            
        \end{tikzpicture}
        &
        \begin{tikzpicture}
            \node[anchor=south west,inner sep=0] (image) {\includegraphics[width=0.16\textwidth]{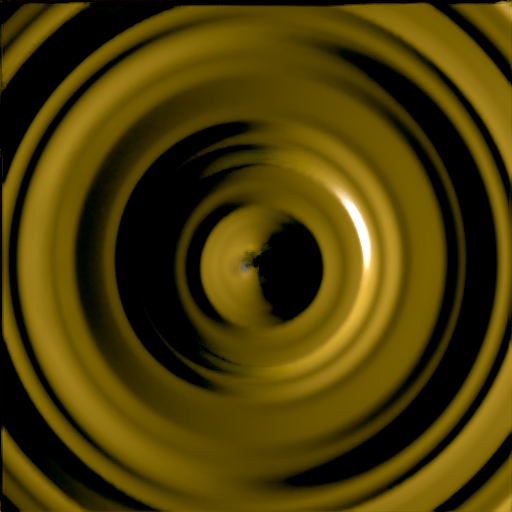}};
            \def\distanceTop{1.5cm};
            \def\distanceBottom{0.5cm};
            \def\distanceLeft{0.5cm};
            \def\distanceRight{1.5cm};
            \draw[red, thick] ([shift={(\distanceLeft,\distanceBottom)}]image.south west) rectangle ([shift={(-\distanceRight,-\distanceTop)}]image.north east);           
        \end{tikzpicture}
        &
        \begin{tikzpicture}
        \node[anchor=south west,inner sep=0] (image)
        {\includegraphics[width=0.16\textwidth]{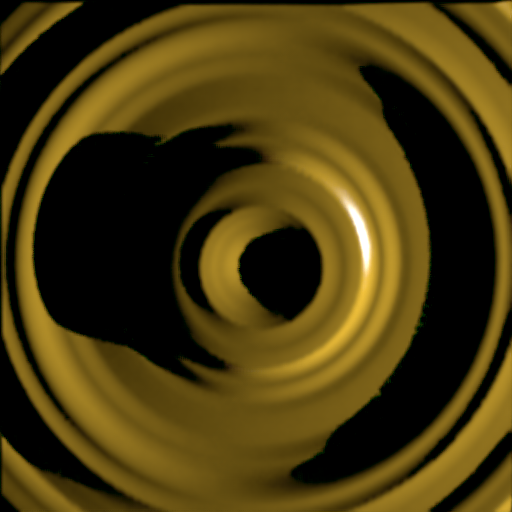}};
            \def\distanceTop{1.5cm};
            \def\distanceBottom{0.5cm};
            \def\distanceLeft{0.5cm};
            \def\distanceRight{1.5cm};
            \draw[red, thick] ([shift={(\distanceLeft,\distanceBottom)}]image.south west) rectangle ([shift={(-\distanceRight,-\distanceTop)}]image.north east);
        \end{tikzpicture}
        &
        \begin{tikzpicture}
            \node[anchor=south west,inner sep=0] (image) {\includegraphics[width=0.16\textwidth]{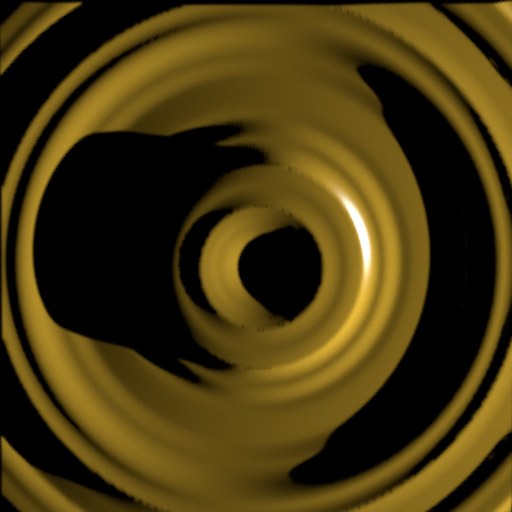}};            
            \def\distanceTop{1.5cm};
            \def\distanceBottom{0.5cm};
            \def\distanceLeft{0.5cm};
            \def\distanceRight{1.5cm};
            \draw[red, thick] ([shift={(\distanceLeft,\distanceBottom)}]image.south west) rectangle ([shift={(-\distanceRight,-\distanceTop)}]image.north east);
        \end{tikzpicture}             
        \\
        \begin{tikzpicture}
        

            \node[anchor=south west,inner sep=0,draw=red,thick] (image) {\includegraphics[width=0.16\textwidth, trim={3cm 3cm 10cm 10cm},clip]{images/ablation2/metalinceptionoff478.png}};
        \end{tikzpicture}
        &
        \begin{tikzpicture}
            \node[anchor=south west,inner sep=0,draw=red,thick] (image) {\includegraphics[width=0.16\textwidth, trim={3cm 3cm 10cm 10cm},clip]{images/ablation2/metalgradoff478.png}};
            
        \end{tikzpicture}
        &
        \begin{tikzpicture}
            \node[anchor=south west,inner sep=0,draw=red,thick] (image) {\includegraphics[width=0.16\textwidth, trim={3cm 3cm 10cm 10cm},clip]{images/ablation2/metalpeoff2478.png}};
            
        \end{tikzpicture}
        &
        \begin{tikzpicture}
            \node[anchor=south west,inner sep=0,draw=red,thick] (image) {\includegraphics[width=0.16\textwidth, trim={3cm 3cm 10cm 10cm},clip]{images/ablation2/metalremapoff478.png}};
           
        \end{tikzpicture}
        &
        \begin{tikzpicture}
            \node[anchor=south west,inner sep=0,draw=red,thick] (image)
            {\includegraphics[width=0.16\textwidth, trim={3cm 3cm 10cm 10cm},clip]{images/newresult2/metalour478.png}};
        \end{tikzpicture}
        &
        \begin{tikzpicture}
            \node[anchor=south west,inner sep=0,draw=red,thick] (image) {\includegraphics[width=0.16\textwidth, trim={3cm 3cm 10cm 10cm},clip]{images/ablation2/gmetalinceptionoff478.png}};
            
        \end{tikzpicture}     
        \\
        \begin{tikzpicture}
            \node[anchor=south west,inner sep=0] (image) {\includegraphics[width=0.16\textwidth]{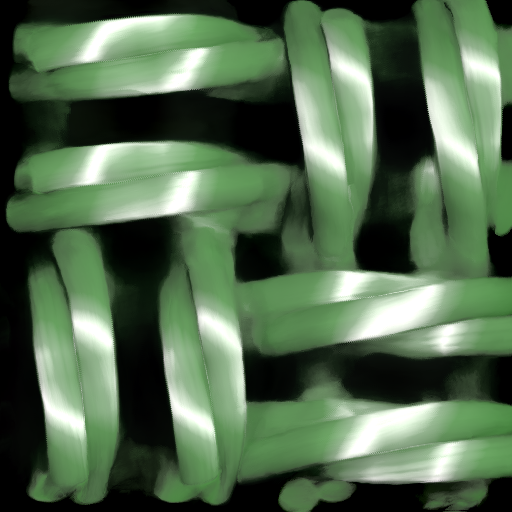}};
            \def\distanceTop{1.7cm};
            \def\distanceBottom{0.3cm};
            \def\distanceLeft{2.0cm};
            \def\distanceRight{0cm};
            \draw[red, thick] ([shift={(\distanceLeft,\distanceBottom)}]image.south west) rectangle ([shift={(-\distanceRight,-\distanceTop)}]image.north east);
        \end{tikzpicture}
        &
        \begin{tikzpicture}
            \node[anchor=south west,inner sep=0] (image) {\includegraphics[width=0.16\textwidth]{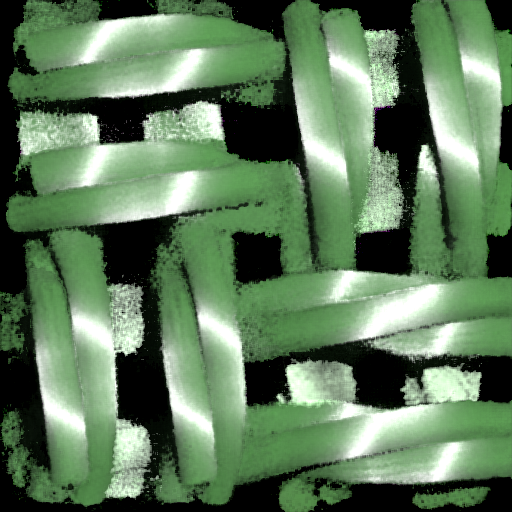}};
            \def\distanceTop{1.7cm};
            \def\distanceBottom{0.3cm};
            \def\distanceLeft{2.0cm};
            \def\distanceRight{0cm};
            \draw[red, thick] ([shift={(\distanceLeft,\distanceBottom)}]image.south west) rectangle ([shift={(-\distanceRight,-\distanceTop)}]image.north east);
        \end{tikzpicture}
        &
        \begin{tikzpicture}
            \node[anchor=south west,inner sep=0] (image) {\includegraphics[width=0.16\textwidth]{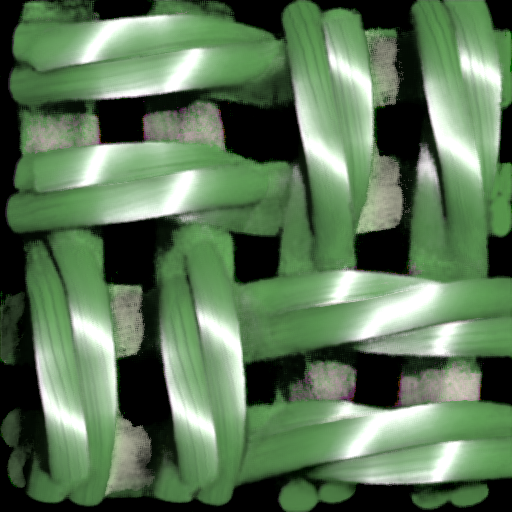}};
            \def\distanceTop{1.7cm};
            \def\distanceBottom{0.3cm};
            \def\distanceLeft{2.0cm};
            \def\distanceRight{0cm};
            \draw[red, thick] ([shift={(\distanceLeft,\distanceBottom)}]image.south west) rectangle ([shift={(-\distanceRight,-\distanceTop)}]image.north east);
        \end{tikzpicture}
        &
        \begin{tikzpicture}
            \node[anchor=south west,inner sep=0] (image) {\includegraphics[width=0.16\textwidth]{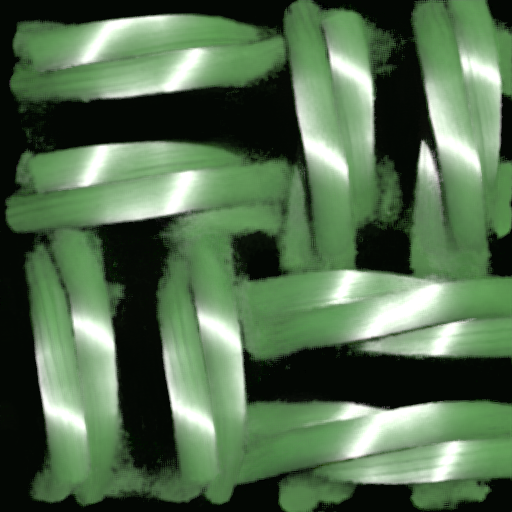}};
            \def\distanceTop{1.7cm};
            \def\distanceBottom{0.3cm};
            \def\distanceLeft{2.0cm};
            \def\distanceRight{0cm};
            \draw[red, thick] ([shift={(\distanceLeft,\distanceBottom)}]image.south west) rectangle ([shift={(-\distanceRight,-\distanceTop)}]image.north east);
        \end{tikzpicture}
        &
        \begin{tikzpicture}
            \node[anchor=south west,inner sep=0] (image)
            {\includegraphics[width=0.16\textwidth]{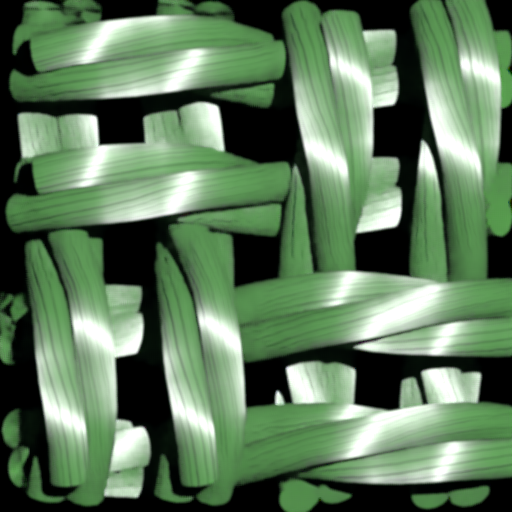}};
            \def\distanceTop{1.7cm};
            \def\distanceBottom{0.3cm};
            \def\distanceLeft{2.0cm};
            \def\distanceRight{0cm};
            \draw[red, thick] ([shift={(\distanceLeft,\distanceBottom)}]image.south west) rectangle ([shift={(-\distanceRight,-\distanceTop)}]image.north east);
        \end{tikzpicture}
        &
        \begin{tikzpicture}
            \node[anchor=south west,inner sep=0] (image) {\includegraphics[width=0.16\textwidth]{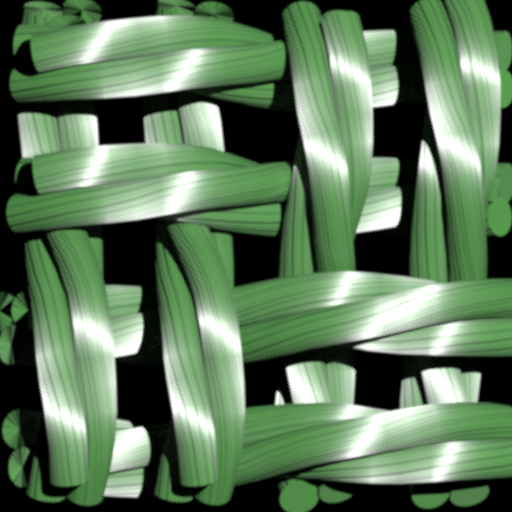}};
            \def\distanceTop{1.7cm};
            \def\distanceBottom{0.3cm};
            \def\distanceLeft{2.0cm};
            \def\distanceRight{0cm};
            \draw[red, thick] ([shift={(\distanceLeft,\distanceBottom)}]image.south west) rectangle ([shift={(-\distanceRight,-\distanceTop)}]image.north east);            
        \end{tikzpicture}
        \\
            \begin{tikzpicture}
        

            \node[anchor=south west,inner sep=0,draw=red,thick] (image) {\includegraphics[width=0.16\textwidth, trim={13cm 2cm 0cm 11cm},clip]{images/ablation2/incepoffgreen4.png}};
        \end{tikzpicture}
        &
        \begin{tikzpicture}
            \node[anchor=south west,inner sep=0,draw=red,thick] (image) {\includegraphics[width=0.16\textwidth, trim={13cm 2cm 0cm 11cm},clip]{images/ablation2/gradientoffgreen4.png}};
            
        \end{tikzpicture}
        &
        \begin{tikzpicture}
            \node[anchor=south west,inner sep=0,draw=red,thick] (image) {\includegraphics[width=0.16\textwidth, trim={13cm 2cm 0cm 11cm},clip]{images/ablation2/peoffgreen4.png}};
            
        \end{tikzpicture}
        &
        \begin{tikzpicture}
            \node[anchor=south west,inner sep=0,draw=red,thick] (image) {\includegraphics[width=0.16\textwidth, trim={13cm 2cm 0cm 11cm},clip]{images/ablation2/remapoffgreen4.png}};
           
        \end{tikzpicture}
        &
        \begin{tikzpicture}
            \node[anchor=south west,inner sep=0,draw=red,thick] (image)
            {\includegraphics[width=0.16\textwidth, trim={13cm 2cm 0cm 11cm},clip]{images/ablation2/ourgreen4.png}};
        \end{tikzpicture}
        &
        \begin{tikzpicture}
            \node[anchor=south west,inner sep=0,draw=red,thick] (image) {\includegraphics[width=0.16\textwidth, trim={13cm 2cm 0cm 11cm},clip]{images/ablation2/gourgreen4.png}};
            
        \end{tikzpicture}   
    \end{tabular}    
    
    \caption{The ablation study was conducted by sequentially deactivating each feature at a time to showcase the effect of each component.}
    \label{fig_ablation}
\end{figure*}

\begin{figure*}[t]
    \setlength{\tabcolsep}{1pt}
    \hspace{-18pt}
    \begin{tabular}{cccccc}
        Orignal size NeuMIP  
        & 10 $\times$ network size  
        & 100 $\times$ network size
        & 300 $\times$ network size 
        & Reference
        & Shape of bump
    
        \\
        \begin{tikzpicture}[baseline]
            \node[anchor=south west,inner sep=0] (image) {\includegraphics[width=0.17\textwidth]{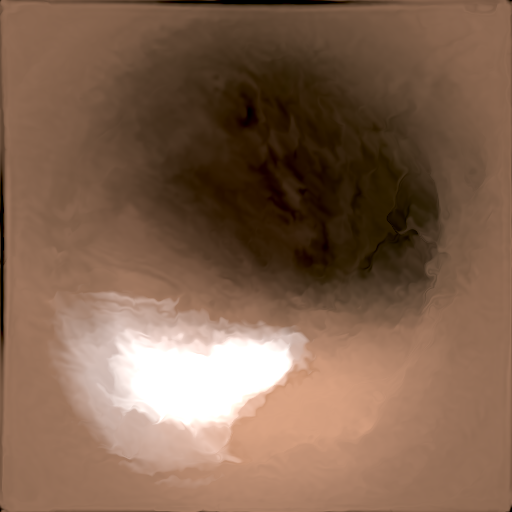}};
            \begin{scope}[x={(image.south east)},y={(image.north west)}]
                \node[anchor=south east, font=\large] at (1,0) {MSE: 1.498}; 
            \end{scope}
        \end{tikzpicture}
        &
        \begin{tikzpicture}[baseline]
            \node[anchor=south west,inner sep=0] (image) {\includegraphics[width=0.17\textwidth]{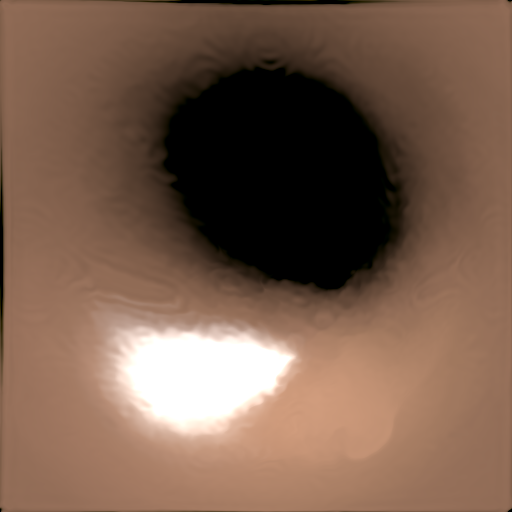}};
            \begin{scope}[x={(image.south east)},y={(image.north west)}]
                \node[anchor=south east, font=\large] at (1,0) {MSE: 0.839};
            \end{scope}
        \end{tikzpicture}
        &
        \begin{tikzpicture}[baseline]
            \node[anchor=south west,inner sep=0] (image) {\includegraphics[width=0.17\textwidth]{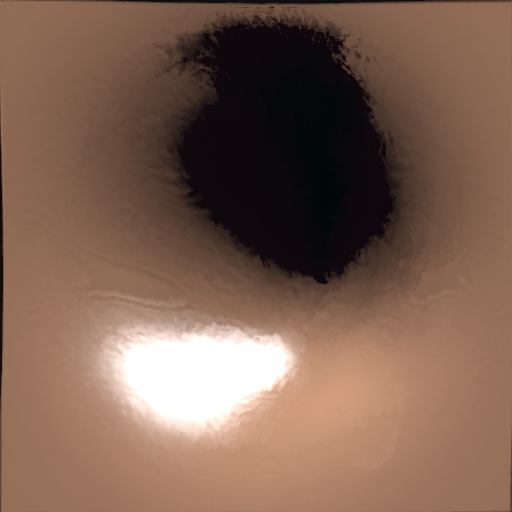}};
            \begin{scope}[x={(image.south east)},y={(image.north west)}]
                \node[anchor=south east, font=\large] at (1,0) {MSE: 0.379}; 
            \end{scope}
        \end{tikzpicture}
        &
        \begin{tikzpicture}[baseline]
            \node[anchor=south west,inner sep=0] (image) {\includegraphics[width=0.17\textwidth]{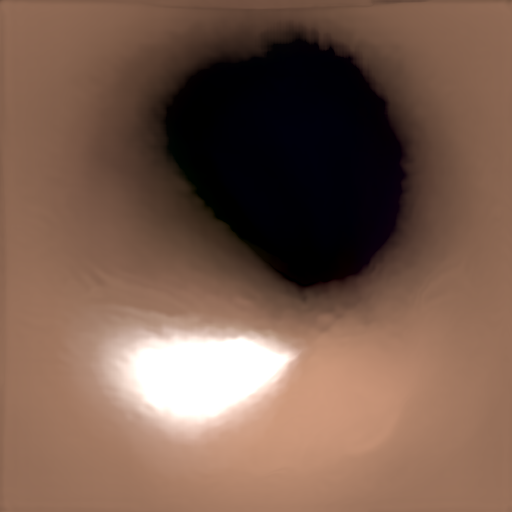}};
            \begin{scope}[x={(image.south east)},y={(image.north west)}]
                \node[anchor=south east, font=\large] at (1,0) {MSE: 0.274};
            \end{scope}
        \end{tikzpicture}
        &
        \includegraphics[width=0.17\textwidth]{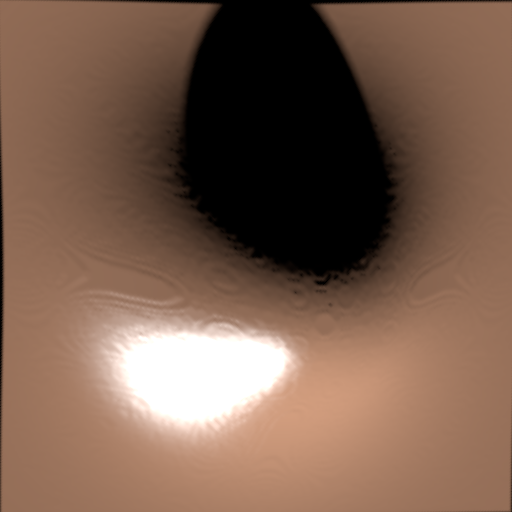}
        &
        \includegraphics[width=0.15\textwidth]{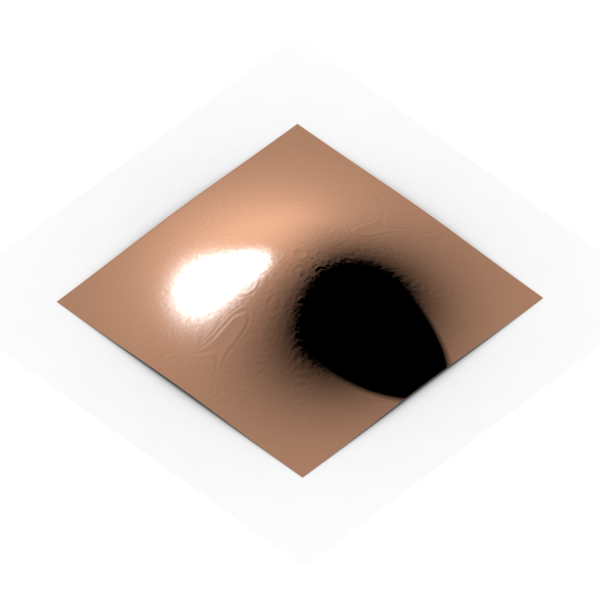}
    \end{tabular}    
    \caption{ We extended NeuMIP by increasing the size of the network and show the larger set of neurons, while being improved as the network extended, still have difficulty capturing the self-shadowing in comparison to the reference. The results are taken after full convergence and noticeable time for training, yet do not fully match the reference.}
    \label{fig_largeneumip}
\end{figure*}

\mypara{Input encodings}
To further improve the effectiveness of our method in handling detailed appearances, we adopt the position encoding that was originally introduced by NeRF\cite{mildenhall2020nerf}.
Specifically, we incorporate high-frequency encoding for lighting direction $\omega_i$ and camera directions $\omega_o$, along with texture position $u$. Rahaman et al. \cite{rahaman2019} showed that neural networks are biased toward learning low-frequency functions and perform poorly at representing high-frequency variation. Thus, we modify the MLP decoder by mapping its inputs to a high-dimensional space using Fourier transformation \cite{fourier} instead of directly operating on input coordinates, such as in previous work. The Fourier transformations are applied to the inputs (wi, wo,u), mapping them to the frequency domain.

This mapping significantly improves the ability of the network to reconstruct highlights and capture high-frequency image features, addressing the shortcomings of the original NeuMIP network. The formulation of our decoder $\decoder$ is a composition of two functions $\decoder = \decoder^\prime \circ \gamma$. Where only $\decoder^\prime$ is learned and $\gamma(.)$ is the encoding function that is applied to each of the input values which are all normalized, $p \in {u, \omega_i, \omega_o}$
\begin{align*}
    \small
    \gamma(p) = \left(sin(2^0 \pi p), cos(2^0 \pi p), ..., sin(2^{L-1} \pi p), cos(2^{L-1} \pi p) \right),
\end{align*}
where L defines the level of frequencies. Based on our experiments, we set L as 10 and 4 for $\gamma(u)$ and $\gamma(\omega)$, respectively. 

The contribution of our proposed architecture capturing the fine details is exhibited in the first two columns of \figref{fig_ablation}.

\subsection{Enhanced Loss Functions}
\label{ssec_loss}
While our hierarchical network design introduced in \S\ref{ssec_network} is crucial for accurately reproducing material appearance at varying scales, training the network using standard image losses (e.g., L1 or L2)
\FloatBarrier
\begin{figure*}[h!]
    \centering
    \newlength{\resLen}
    \setlength{\resLen}{.2\textwidth}
    \newlength{\insetLen}
    \setlength{\insetLen}{.099\textwidth}
    \hspace{-18pt}
    \vspace{-10pt}
    \setlength{\tabcolsep}{1.5pt}
    \begin{tabular}{ccccc}
        & {Reference}
        & {Ours}
        & {NeuMIP}
        & Reference  \quad \qquad Ours \qquad \quad NeuMIP
        \\ 
        
        \raisebox{30pt}{\rotatebox{90}{a) Metal ring}} &
        \includegraphics[width=\resLen]{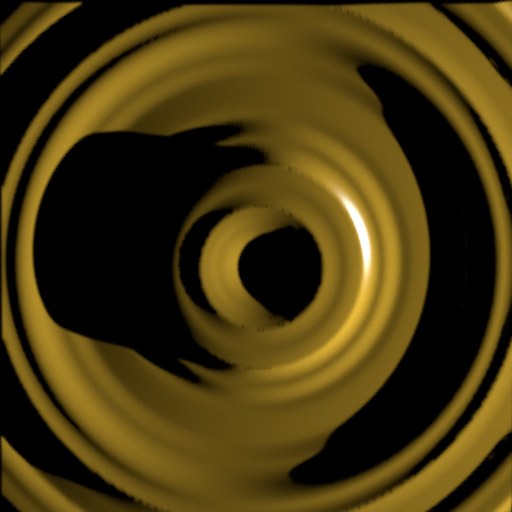}&
        \includegraphics[width=\resLen]{images/newresult2/metalour478.png}&
        \includegraphics[width=\resLen]{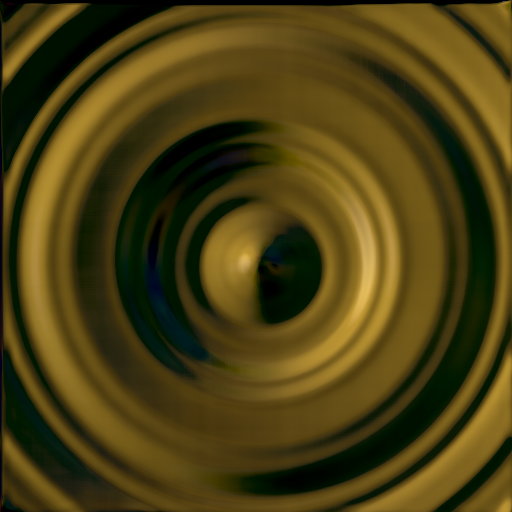}&
        \begin{tikzpicture}
        \draw (0, 1.8) node[inner sep=0] {\includegraphics[width=\insetLen, trim={1cm 9cm 11cm 3cm},clip]{images/newresult2/gmetalour478.png}};
        \draw (1.8,1.8) node[inner sep=0] {\includegraphics[width=\insetLen, trim={1cm 9cm 11cm 3cm},clip]{images/newresult2/metalour478.png}};
        \draw (3.6,1.8) node[inner sep=0] {\includegraphics[width=\insetLen, trim={1cm 9cm 11cm 3cm},clip]{images/newresult2/metalneu478.png}};
        \draw (0, 0) node[inner sep=0] {\includegraphics[width=\insetLen, trim={11cm 6cm 1cm 6cm},clip]{images/newresult2/gmetalour478.png}};
        \draw (1.8,0) node[inner sep=0] {\includegraphics[width=\insetLen, trim={11cm 6cm 1cm 6cm},clip]{images/newresult2/metalour478.png}};
        \draw (3.6, 0) node[inner sep=0] {\includegraphics[width=\insetLen, trim={11cm 6cm 1cm 6cm},clip]{images/newresult2/metalneu478.png}};
        \end{tikzpicture}  
        \\
        \raisebox{30pt}{\rotatebox{90}{b) UBO Leather}} &
        \includegraphics[width=\resLen]{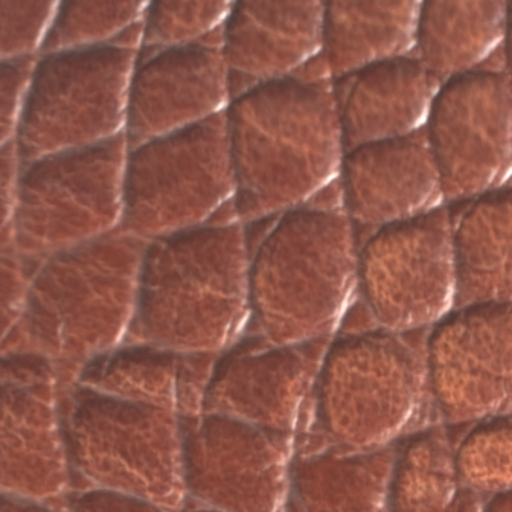}&
        \includegraphics[width=\resLen]{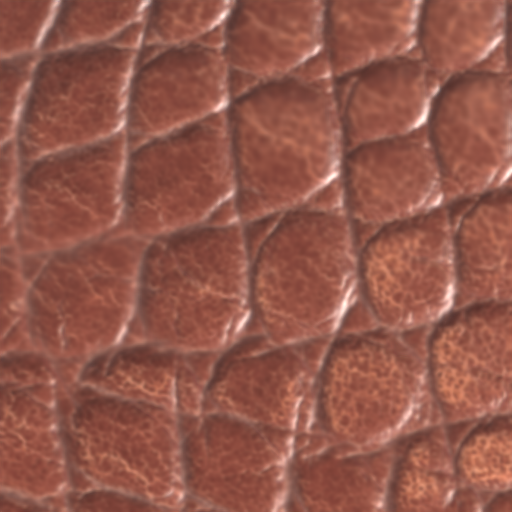}&
         \includegraphics[width=\resLen]{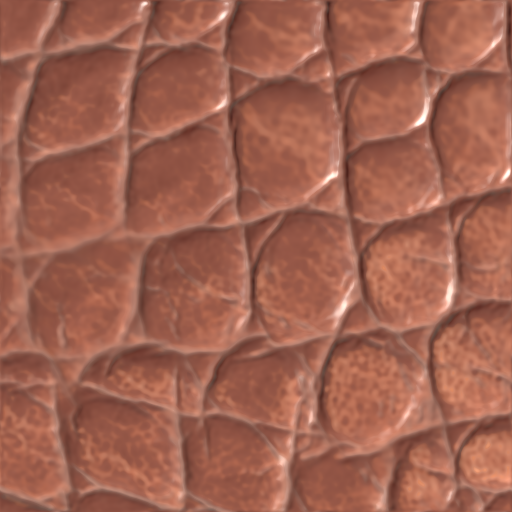}&
        \begin{tikzpicture}
        \draw (0, 1.8) node[inner sep=0] {\includegraphics[width=\insetLen, trim={1cm 4cm 11cm 8cm},clip]{images/gleatherour144.png}};
        \draw (1.8,1.8) node[inner sep=0] {\includegraphics[width=\insetLen, trim={1cm 4cm 11cm 8cm},clip]{images/leatherour144.png}};
        \draw (3.6,1.8) node[inner sep=0] {\includegraphics[width=\insetLen, trim={1cm 4cm 11cm 8cm},clip]{images/leatherneu144.png}};
        \draw (0, 0) node[inner sep=0] {\includegraphics[width=\insetLen, trim={0cm 12cm 12cm 0cm},clip]{images/gleatherour144.png}};
        \draw (1.8,0) node[inner sep=0] {\includegraphics[width=\insetLen, trim={0cm 12cm 12cm 0cm},clip]{images/leatherour144.png}};
        \draw (3.6, 0) node[inner sep=0] {\includegraphics[width=\insetLen, trim={0cm 12cm 12cm 0cm},clip]{images/leatherneu144.png}};
        \end{tikzpicture} 
        \\

        \raisebox{30pt}{\rotatebox{90}{c) Metal grid}} &
        \includegraphics[width=\resLen]{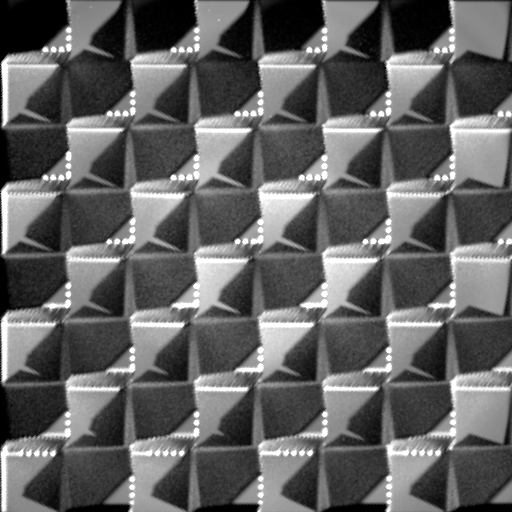}&
        \includegraphics[width=\resLen]{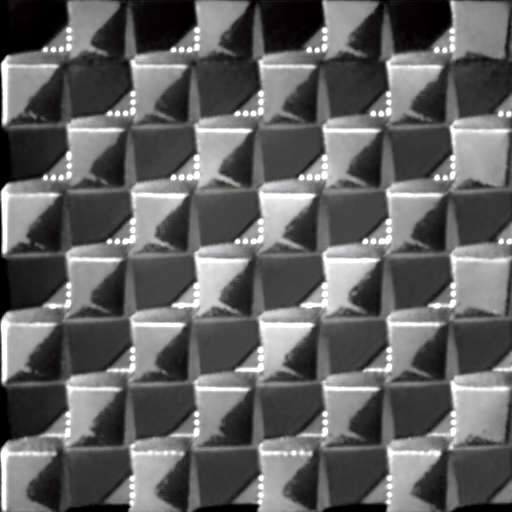}&
         \includegraphics[width=\resLen]{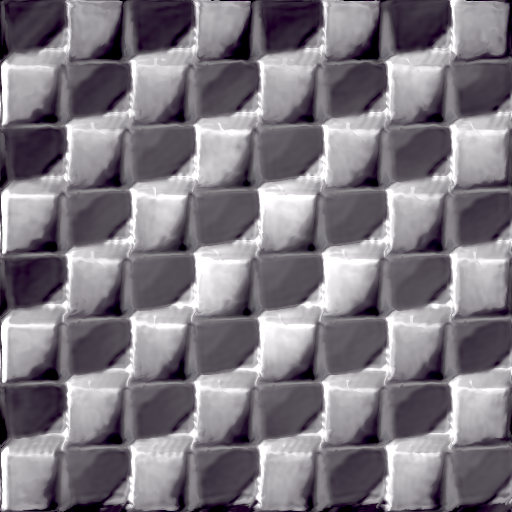}&
        \begin{tikzpicture}
        \draw (0, 1.8) node[inner sep=0] {\includegraphics[width=\insetLen, trim={7cm 6cm 9cm 10cm},clip]{images/newresult2/gmg9ourfull}};
        \draw (1.8,1.8) node[inner sep=0] {\includegraphics[width=\insetLen, trim={7cm 6cm 9cm 10cm},clip]{images/newresult2/mg9ourfull}};
        \draw (3.6,1.8) node[inner sep=0] {\includegraphics[width=\insetLen, trim={7cm 6cm 9cm 10cm},clip]{images/newresult2/mgneu9}};
        \draw (0, 0) node[inner sep=0] {\includegraphics[width=\insetLen, trim={3cm 9cm 13cm 7cm},clip]{images/newresult2/gmg9ourfull}};
        \draw (1.8,0) node[inner sep=0] {\includegraphics[width=\insetLen, trim={3cm 9cm 13cm 7cm},clip]{images/newresult2/mg9ourfull}};
        \draw (3.6, 0) node[inner sep=0] {\includegraphics[width=\insetLen, trim={3cm 9cm 13cm 7cm},clip]{images/newresult2/mgneu9}};
        \end{tikzpicture}

        \\
        \raisebox{30pt}{\rotatebox{90}{d) Turtle shell}} &
        \includegraphics[width=\resLen, trim={2.5cm 0cm 0cm 2.5cm},clip]{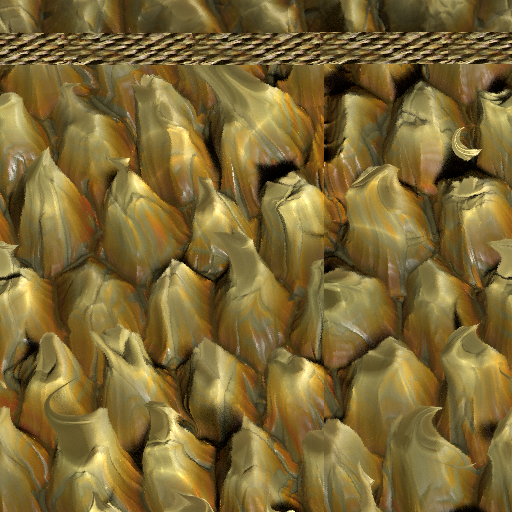}&
        \includegraphics[width=\resLen, trim={2.5cm 0cm 0cm 2.5cm},clip]{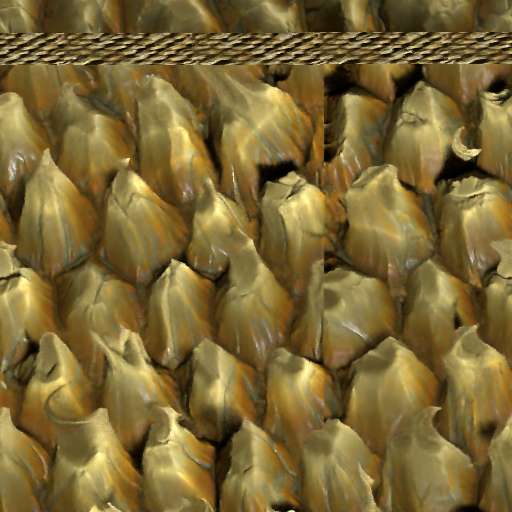}&
        \includegraphics[width=\resLen, trim={2.5cm 0cm 0cm 2.5cm},clip]{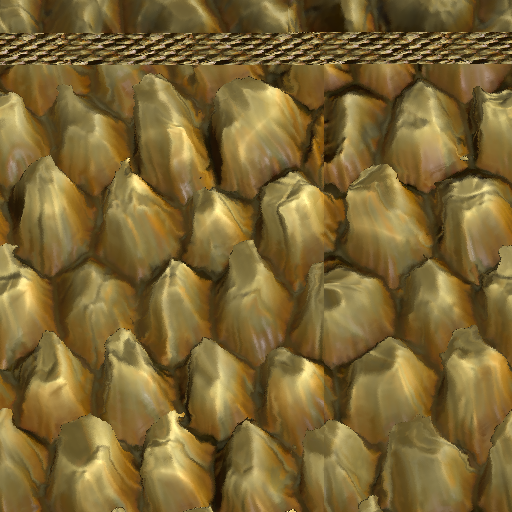}&
        \begin{tikzpicture}
        \draw (0, 1.8) node[inner sep=0] {\includegraphics[width=\insetLen, trim={8cm 10cm 7cm 5cm},clip]{images/newresult2/gshellourfull144.png}};
        \draw (1.8,1.8) node[inner sep=0] {\includegraphics[width=\insetLen, trim={8cm 10cm 7cm 5cm},clip]{images/newresult2/shellourfull144.png}};
        \draw (3.6,1.8) node[inner sep=0] {\includegraphics[width=\insetLen, trim={8cm 10cm 7cm 5cm},clip]{images/newresult2/shellneu144.png}};
        \draw (0, 0) node[inner sep=0] {\includegraphics[width=\insetLen, trim={11cm 4cm 4cm 11cm},clip]{images/newresult2/gshellourfull144.png}};
        \draw (1.8,0) node[inner sep=0] {\includegraphics[width=\insetLen, trim={11cm 4cm 4cm 11cm},clip]{images/newresult2/shellourfull144.png}};
        \draw (3.6, 0) node[inner sep=0] {\includegraphics[width=\insetLen, trim={11cm 4cm 4cm 11cm},clip]{images/newresult2/shellneu144.png}};
        \end{tikzpicture} 
        \\
        \raisebox{20pt}{\rotatebox{90}{e) Victorian fabric}} &
        \includegraphics[width=\resLen]{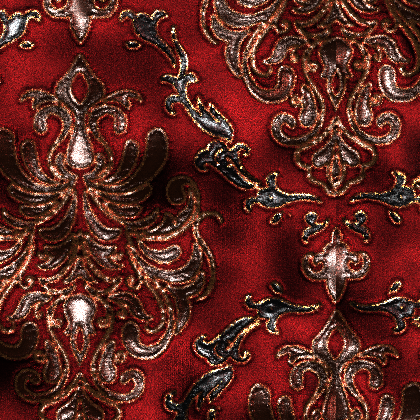}&
        \includegraphics[width=\resLen]{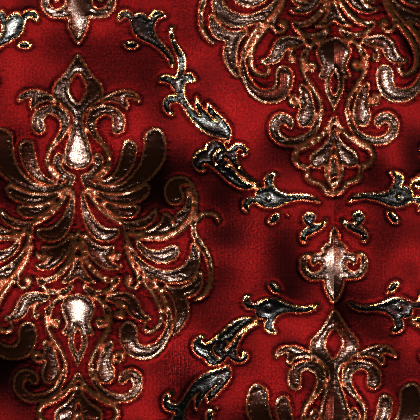}&
        \includegraphics[width=\resLen]{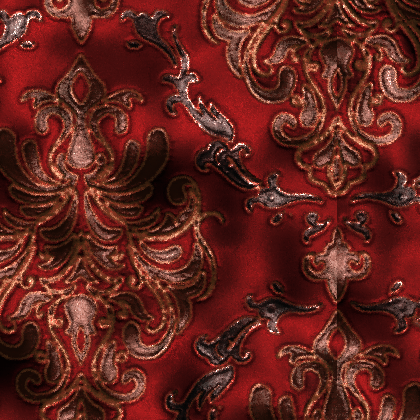}&
       
        \begin{tikzpicture}
        \draw (0, 1.8) node[inner sep=0] {\includegraphics[width=\insetLen, trim={2cm 8cm 10cm 4cm},clip]{images/result/gourvictorian32_cut}};
        \draw (1.8, 1.8) node[inner sep=0] {\includegraphics[width=\insetLen, trim={2cm 8cm 10cm 4cm},clip]{images/result/ourvictorian32_cut}};
        \draw (3.6, 1.8) node[inner sep=0] {\includegraphics[width=\insetLen, trim={2cm 8cm 10cm 4cm},clip]{images/result/victorian32_cut}};
        \draw (0, 0) node[inner sep=0] {\includegraphics[width=\insetLen, trim={10cm 4cm 2cm 8cm},clip]{images/result/gourvictorian32_cut}};
        \draw (1.8,0) node[inner sep=0] {\includegraphics[width=\insetLen, trim={10cm 4cm 2cm 8cm},clip]{images/result/ourvictorian32_cut}};
        \draw (3.6, 0) node[inner sep=0] {\includegraphics[width=\insetLen, trim={10cm 4cm 2cm 8cm},clip]{images/result/victorian32_cut}};
        \end{tikzpicture} 
         \\
         \raisebox{30pt}{\rotatebox{90}{f) Twill cloth}} &
        \includegraphics[width=\resLen]{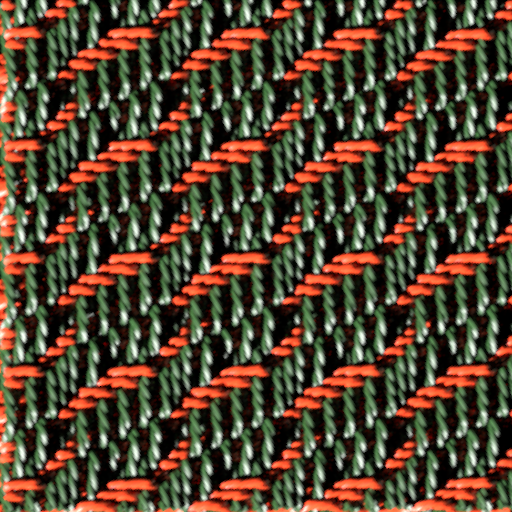}&
        \includegraphics[width=\resLen]{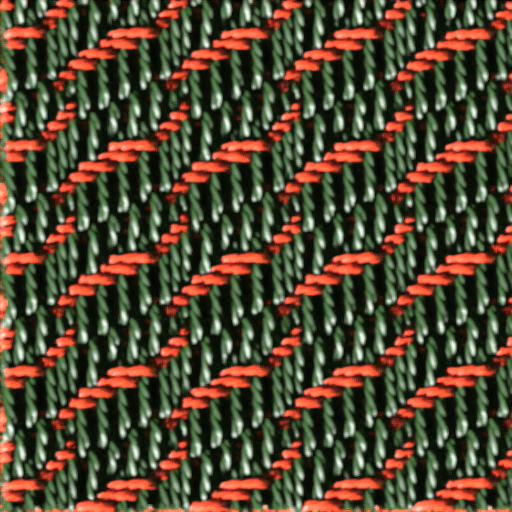}&
         \includegraphics[width=\resLen]{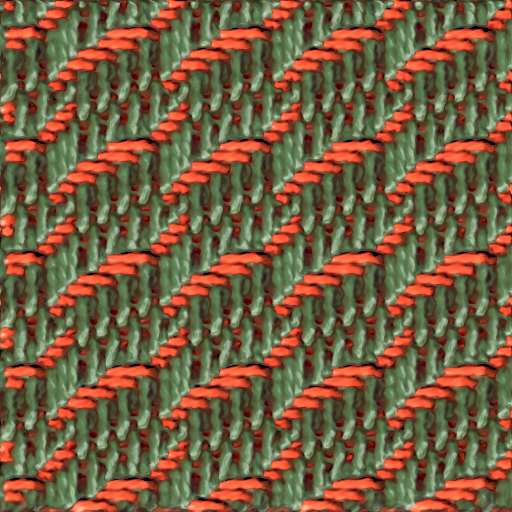}&

        \begin{tikzpicture}
        \draw (0, 1.8) node[inner sep=0] {\includegraphics[width=\insetLen, trim={5cm 6cm 9cm 8cm},clip]{images/newresult2/g2courfull5.png}};
        \draw (1.8,1.8) node[inner sep=0] {\includegraphics[width=\insetLen, trim={5cm 6cm 9cm 8cm},clip]{images/newresult2/2courfull5.png}};
        \draw (3.6,1.8) node[inner sep=0] {\includegraphics[width=\insetLen, trim={5cm 6cm 9cm 8cm},clip]{images/newresult2/2cneu5.png}};
        \draw (0, 0) node[inner sep=0] {\includegraphics[width=\insetLen, trim={12cm 3cm 2cm 11cm},clip]{images/newresult2/g2courfull5.png}};
        \draw (1.8,0) node[inner sep=0] {\includegraphics[width=\insetLen, trim={12cm 3cm 2cm 11cm},clip]{images/newresult2/2courfull5.png}};
        \draw (3.6, 0) node[inner sep=0] {\includegraphics[width=\insetLen, trim={12cm 3cm 2cm 11cm},clip]{images/newresult2/2cneu5.png}};
        \end{tikzpicture}

    \end{tabular}    
    \caption{Comparisons of real and synthetic data with the reference. Please see the accompanying video for further comparisons.  }
    \label{fig_comparison}
\end{figure*}

\FloatBarrier
may lead to results that still lack details. To address this issue, we propose using the following losses for training.


\mypara{Gradient loss.}
Inspired by the Canny edge detection algorithm \cite{canny}, we utilize a gradient loss to encourage the network to better preserve detailed shading variations:
\begin{equation}
    \loss_G (\image; \hat{\image}) = \left( G_x(\image) - G_x(\hat{\image}) \right) ^2 + \left( G_y(\image) - G_y(\hat{\image}) \right)^2,
    \label{eq_loss_G}
\end{equation}

where, for any $\image$, $G_x(\image) := \kernel_x \ast \image$ and $G_y(\image) := \kernel_y \ast \image$ indicate, respectively, the image $\image$ convolved with the Sobel edge-detection filters \cite{sobel}
\begin{align*}
    \kernel_x = 
    \begin{bmatrix}
        1 & 0 & -1\\
        2 & 0 & -2\\
        1 & 0 & -1
    \end{bmatrix} 
    \quad\text{and}\quad
    \kernel_y = 
    \begin{bmatrix}
        1 & 2 & 1\\
        0 & 0 & 0\\
        -1 & -2 & -1
    \end{bmatrix}.
\end{align*}
Additionally, $\hat{I}$ is the reference image. The new gradient loss only works in the training phase and does not influence the network evaluation. 

\mypara{Output remapping}
The perception of an object's luminance by humans is inherently nonlinear~\cite{bertalmio2020evidence}. However, neural networks tend to minimize global loss, equating the same numerical loss in both high and low-luminance regions. This is in contrast to human perception where the numerical loss in low luminance regions is more pronounced. Building upon this insight, we introduce an ``output remapping'' strategy to assist neural network learning. Once the neural network predicts (linear) RGB values, this remapping assigns different weights based on luminance. Specifically, The remapping is applied to the texture (both reference and generate results) to compute the loss during the training phase. Subsequently, these weighted values are passed through our gradient loss Eqn ~\eqref{eq_loss_G}, leading to a significant enhancement in the quality of shadows and darker regions, with no adverse effect on the image's overall brightness. The new gradient loss only works in the training phase, so would not influence the network evaluation. Therefore, our proposed strategy achieves superior results in these two loss metrics.
After rigorous experiments, we noticed applying $4^{th}$ root functions as the image remapping is the optimum spot to capture both low and high frequencies. Our final loss function is formulated as follows:

\begin{align}
 \loss =
  \frac{1}{n}\mathlarger{\sum_{i=0}^n} \left( \loss_1 (\frac{1}{\image^4}) + \loss_G (\frac{1} {\image^4}) \right),
\end{align}
where $\image^{-4}$ and $\hat{\image}^{-4}$ are obtained by applying per-pixel exponents to the output and reference images, and $\loss_G$ is the gradient loss defined in Eqn \eqref{eq_loss_G}.

\begin{table}[bth]
\setlength{\tabcolsep}{2pt}
\caption{Errors for images in \figref{fig_comparison}.}
\begin{tabular}{lcccccc}
\toprule
& \multicolumn{2}{c}{MSE} & \multicolumn{2}{c}{LPIPS} & \multicolumn{2}{c}{PSNR} \\
\cmidrule(lr){2-3}\cmidrule(lr){4-5}\cmidrule(lr){6-7}
Scene $\downarrow$ & Ours & NeuMIP & Ours & NeuMIP & Ours & NeuMIP \\
\midrule
a) Ring & 0.342 & 3.474 & 0.139 & 0.215 & 38.472 & 29.887 \\
b) Leather & 0.029 & 0.540 & 0.012 & 0.134 & 32.792 & 28.601 \\
c) Metal grid & 0.076 & 0.401 & 0.156 & 0.314 & 30.355 & 28.066 \\
d) Turtle shell & 0.056 & 0.703 & 0.070 & 0.163 & 32.181 & 30.785 \\
e) Victorian fabric & 1.335 & 8.071 & 0.104 & 0.141 & 35.252 & 30.849 \\
f) Twill cloth & 0.005 & 0.018 & 0.120 & 0.262 & 29.507 & 28.157 \\
\bottomrule
\end{tabular}

    \label{table_error}
\end{table}
\section{Implementation}
\label{sec_implementation}

\subsection{Dataset and training}
Identical to NeuMIP, our neural appearance model takes as input 7D queries (expressing the camera and light directions $\omega_o$ and $\omega_i$, UV location $u$, and the prefilter kernel size $\sigma$) and outputs a single 3D vector indicating the corresponding RGB reflectance value.
For each material, our training data involves a large set of input-output value pairs.
To train our model, we minimize the loss discussed in \S\ref{ssec_loss} using the Adam algorithm.

In practice, we generate our synthetic training datasets (Basket and Twill cloth, Metal ring, Bump) using the Keyshot path-tracer renderer~\cite{keyshot}. Specifically, the Metal ring and Bump data are rendered using displaced geometry (expressed using height maps). The Basket and Twill cloth datasets, on the other hand, use state-of-the-art ply-based cloth models~\cite{montazeri2020practical, montazeri2021practical}.
Our generated data involves 500 input-output value pairs, and our training process uses mini-batching with a batch size of $30,000$.

Additionally, we use two datasets (Victorian cloth and Turtle shell) published by NeuMIP~\cite{kuznetsov2021neumip} to evaluate our method. We retrained NeuMIP’s victorian fabric and turtle shell datasets with the training parameters tweaked for best results. 

The training of our model as well as the ordinary NeuMIP is performed per material and uses all available training data.
Our training of one material model takes about 90 minutes.

\subsection{Rendering}
Similar to NeuMIP, our neural reflectance model can be integrated into Monte-Carlo renderers. The results shown in this paper use an implementation in Mitsuba 2 rendering engine \cite{Mitsuba}, accounting only for direct illumination. At render time, we use material query buffers (storing $u$, $\omega_i$, $\omega_o$, $sigma$) to compute the inputs to our framework, then we pass the whole buffer to GPU to evaluate the queries as a batch. The LoD is also accounted for in rendered results based on the camera distance per query. All the comparison results have 1 sample-per-pixel (SPP).

\subsection{Performance}
Our model requires approximately 0.035 seconds to generate a 512x512 texture using NVIDIA V100 GPU, compared to around 0.028 seconds for NeuMIP, which is a marginal difference. Given the added complexity of convolution architecture, the number of parameters will increase and the time needed for evaluating our network is about 25\% longer than that of the single-resolution NeuMIP. This increase in evaluation time could be seen as a justifiable trade-off for achieving more accurate quality in capturing details.



        
        

\begin{figure*}[h!]
    \centering
    \setlength{\tabcolsep}{2pt}
    \hspace{-40pt}
    \begin{tabular}{ccc}
        
        \rotatebox{90}{\hspace{-90pt}{NeuMIP}\hspace{45pt}{Ours}\hspace{45pt}{Reference}
        }

        \begin{tabular}{c}
            Basket cloth\\
            \includegraphics[width=0.48\textwidth]{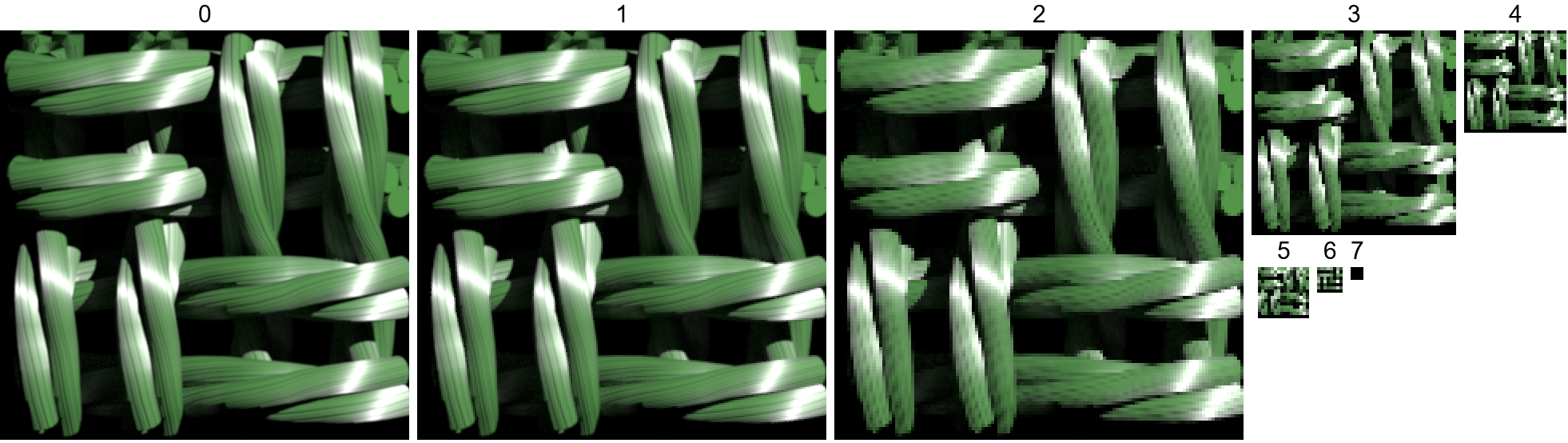} \\
            \includegraphics[width=0.48\textwidth]{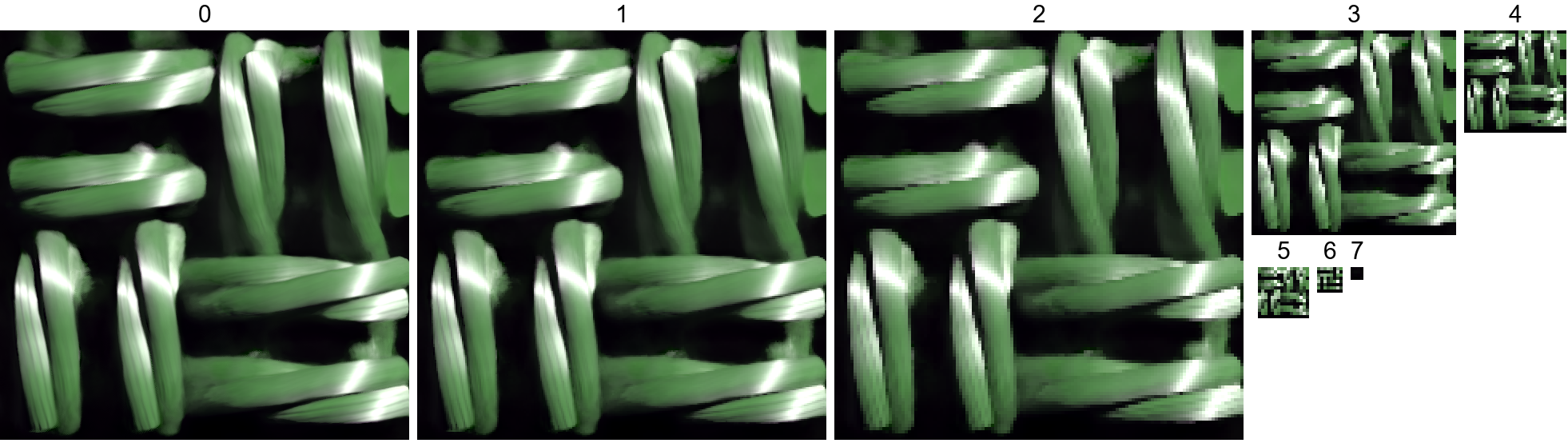} \\
            \includegraphics[width=0.48\textwidth]{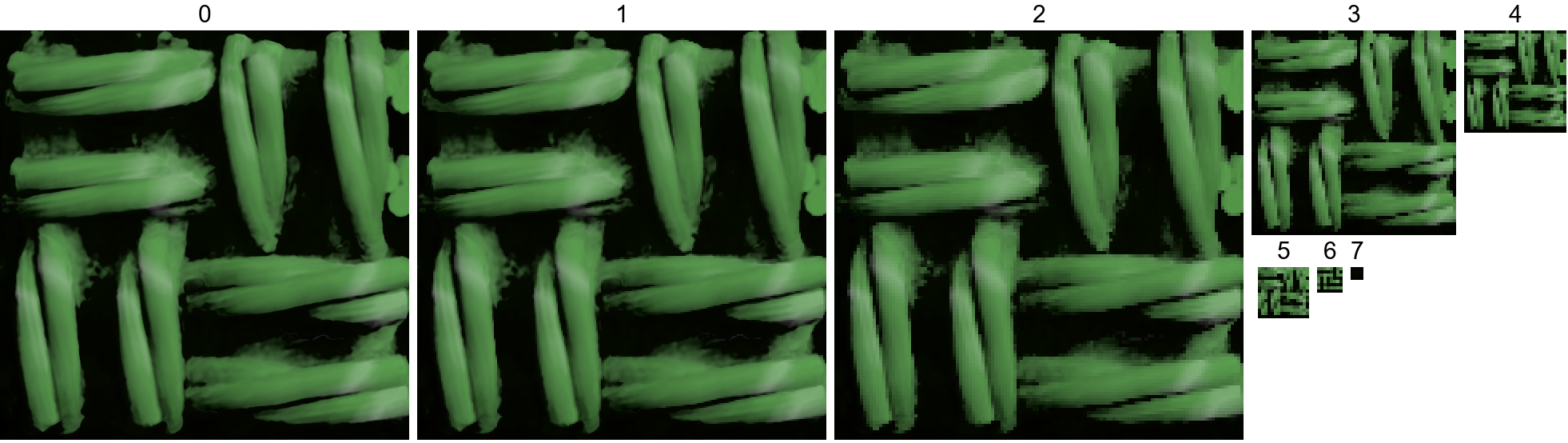}
        \end{tabular} &
        \quad
        &
        \begin{tabular}{c}
        Metal ring\\
            \includegraphics[width=0.48\textwidth]{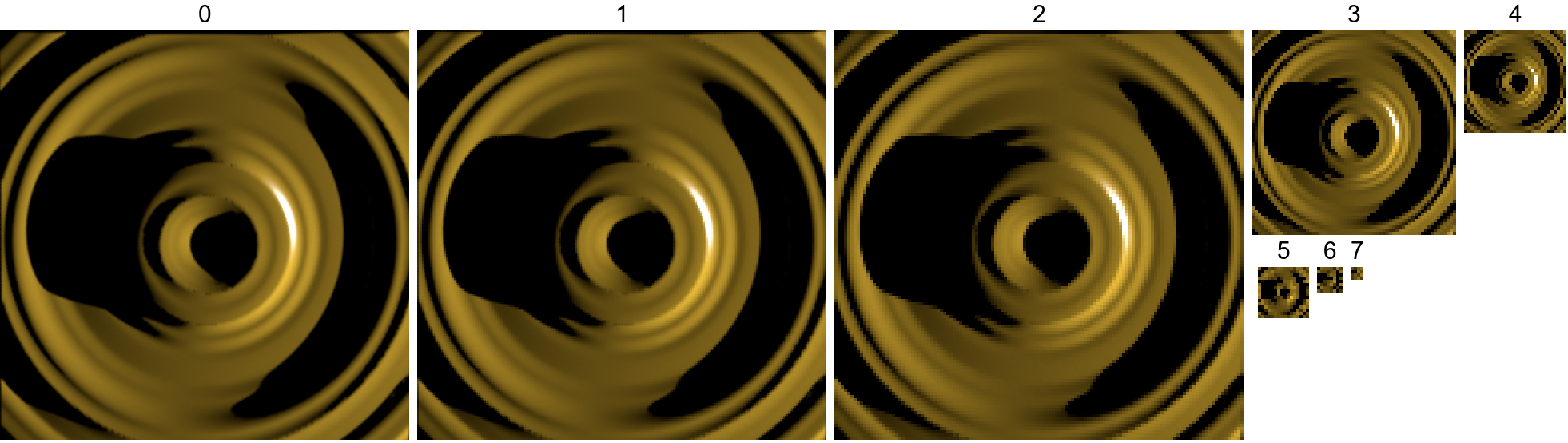} \\
            \includegraphics[width=0.48\textwidth]{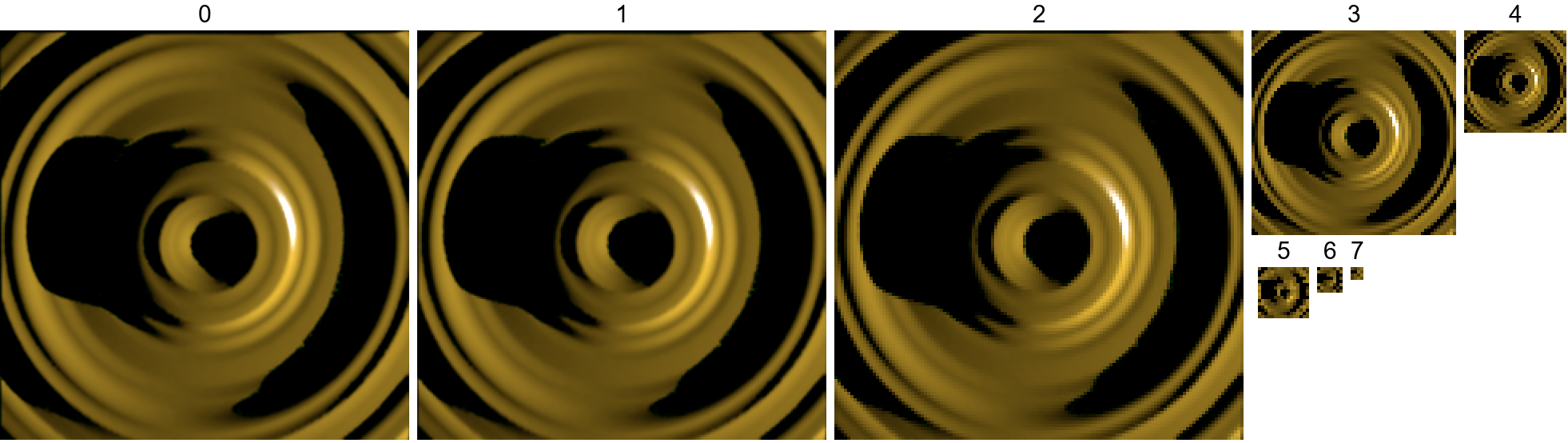} \\
            \includegraphics[width=0.48\textwidth]{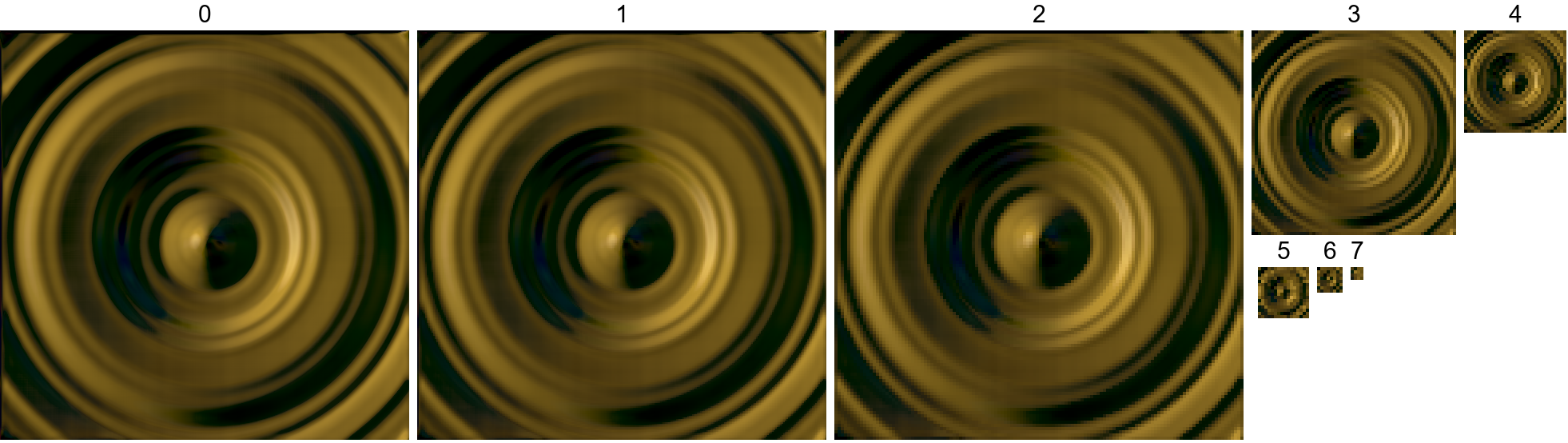}
        \end{tabular}
        
    \end{tabular}

    \caption{Rendered results at the different levels of detail for selected materials.}
    \label{fig_LoD}
\end{figure*}
\begin{figure*}[h!]
    \centering
    \setlength{\tabcolsep}{1pt}
    \hspace{-20pt}
    \begin{tabular}{cccc}
        \includegraphics[width=0.25\textwidth,trim={0 2cm 0 3cm},clip]{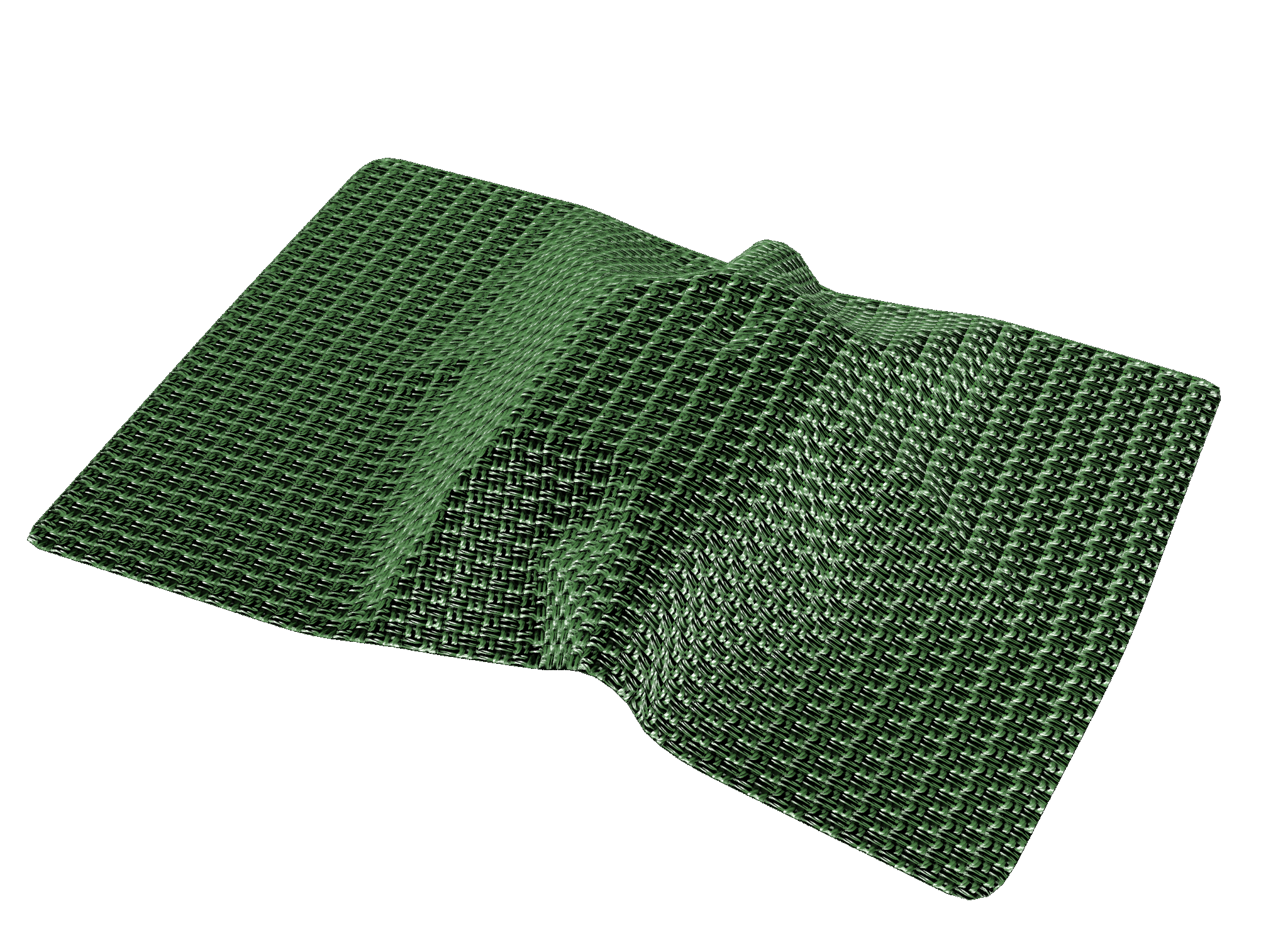}&
        \includegraphics[width=0.25\textwidth,trim={0 0 0 7cm},clip]{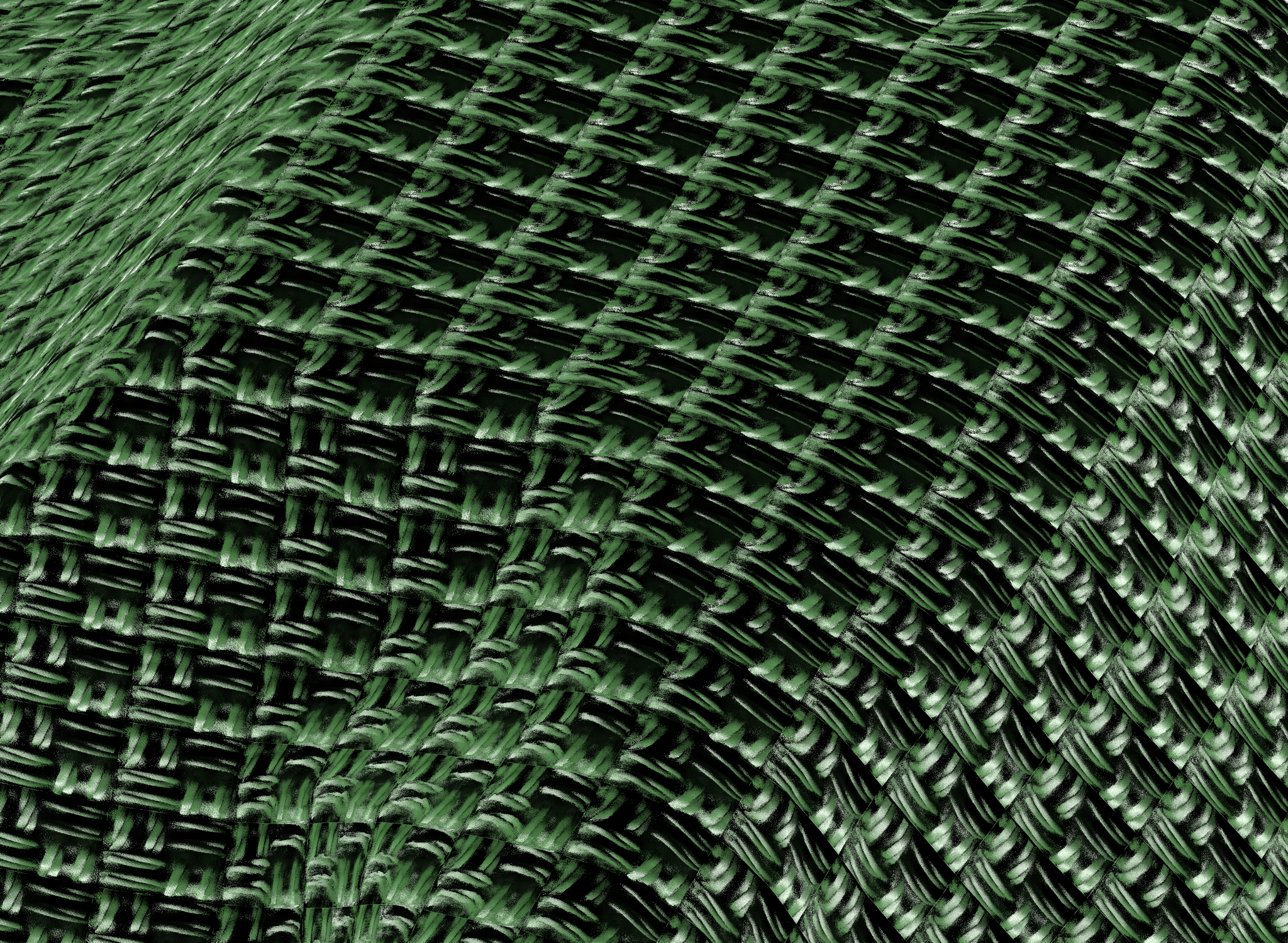}&       
        \includegraphics[width=0.25\textwidth,trim={0 2cm 0 3cm},clip]{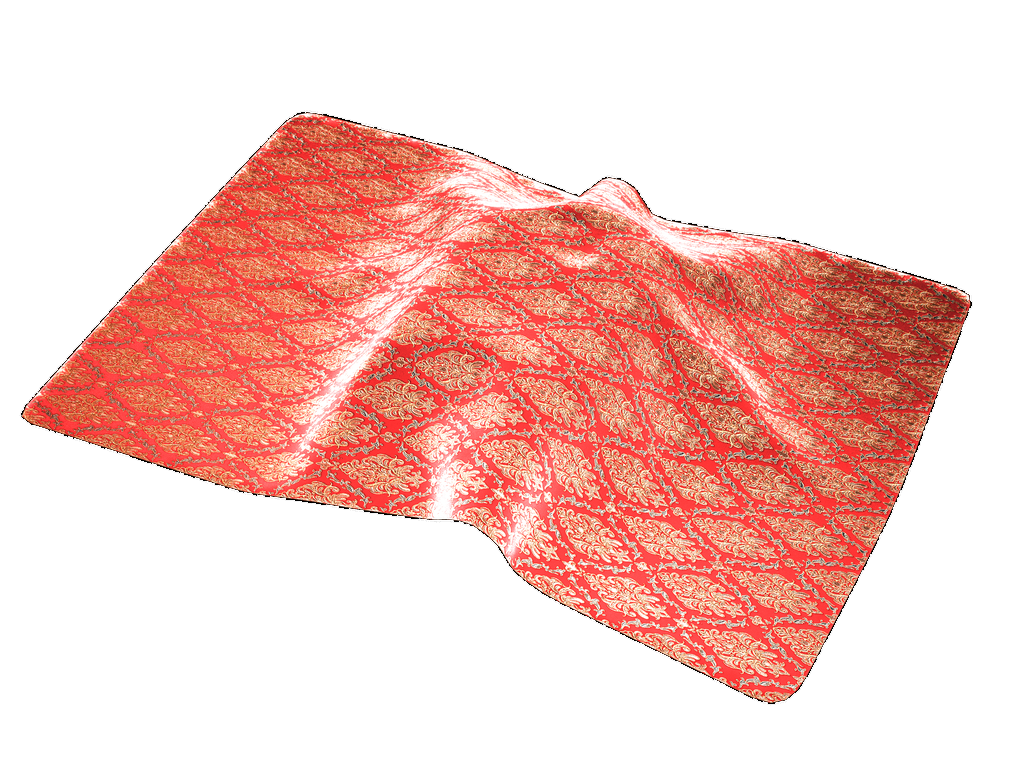}&
        \includegraphics[width=0.25\textwidth,trim={0 0 0 7cm},clip]{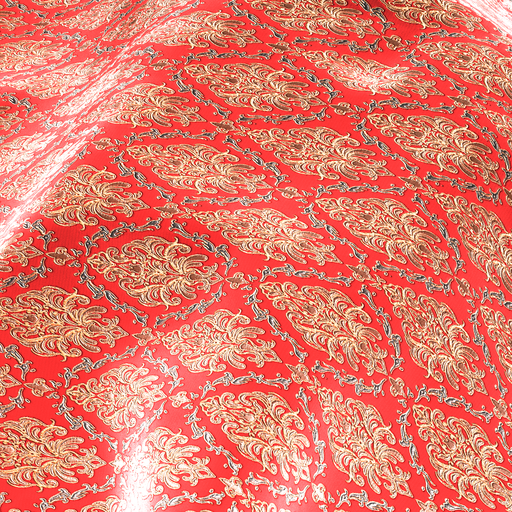}
        \\
        \includegraphics[width=0.25\textwidth,trim={0 2cm 0 3cm},clip]{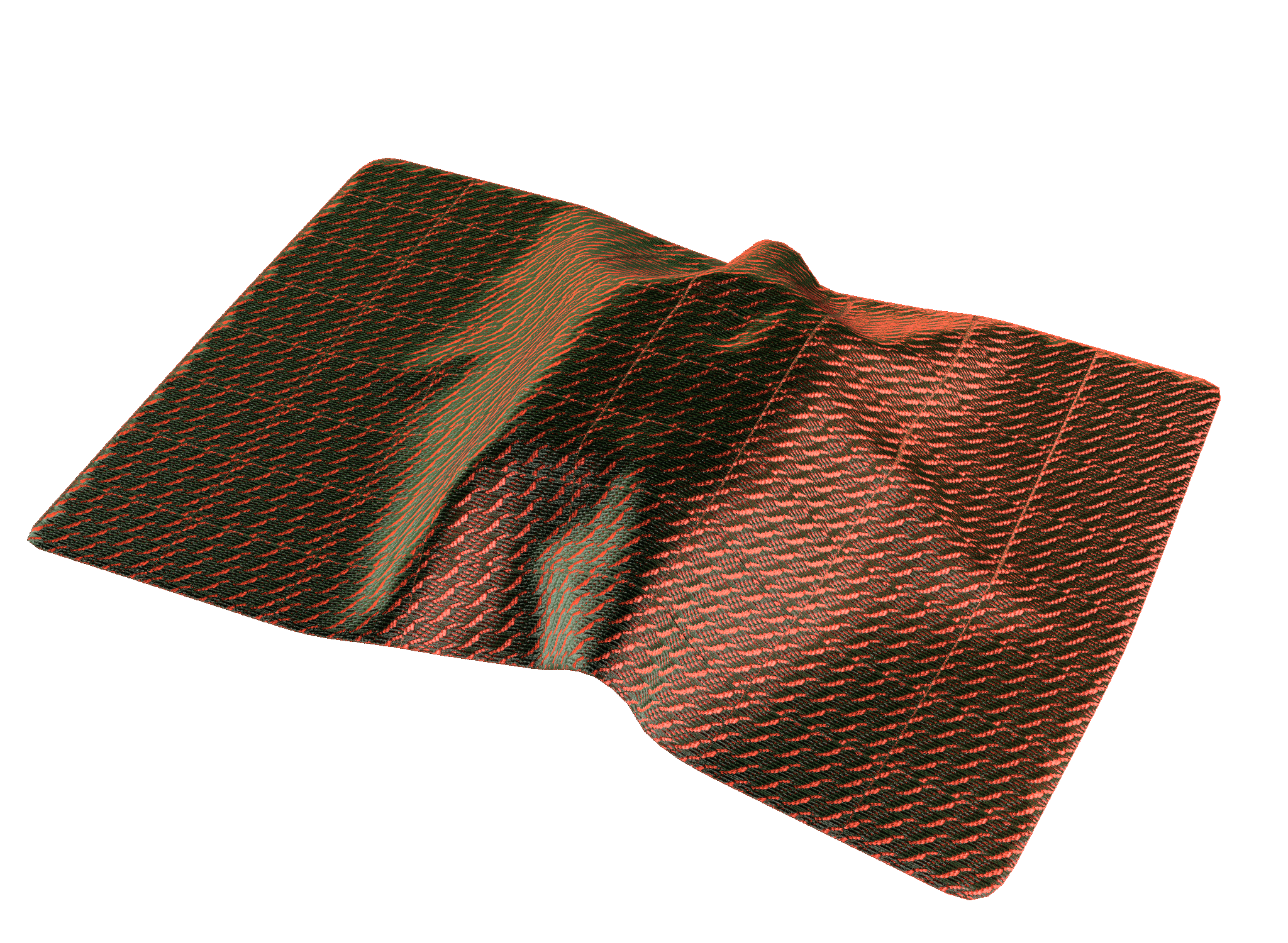}&
        \includegraphics[width=0.25\textwidth,trim={0 0 0 7cm},clip]{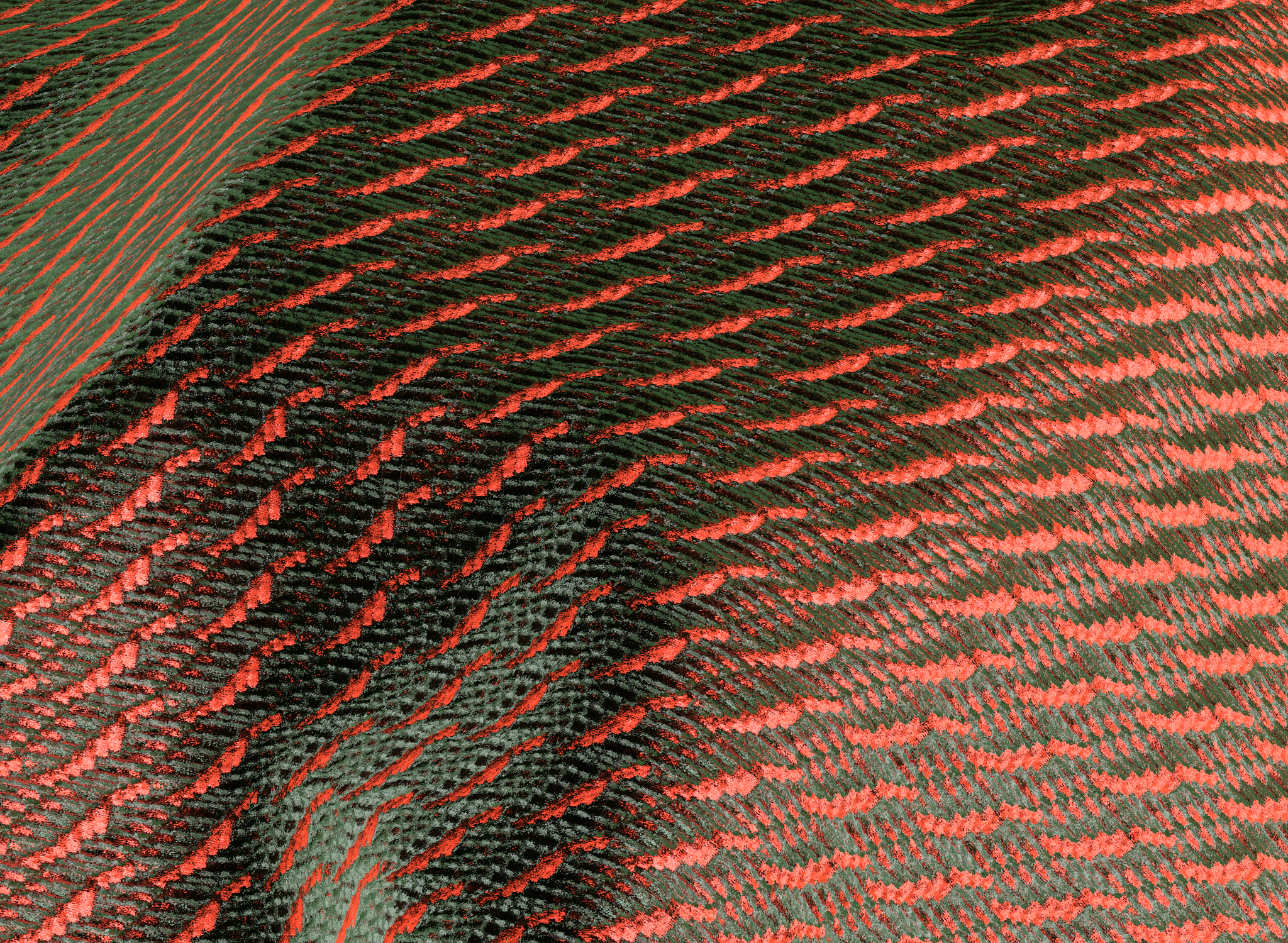}&
         \includegraphics[width=0.25\textwidth,trim={0 2cm 0 3cm},clip]{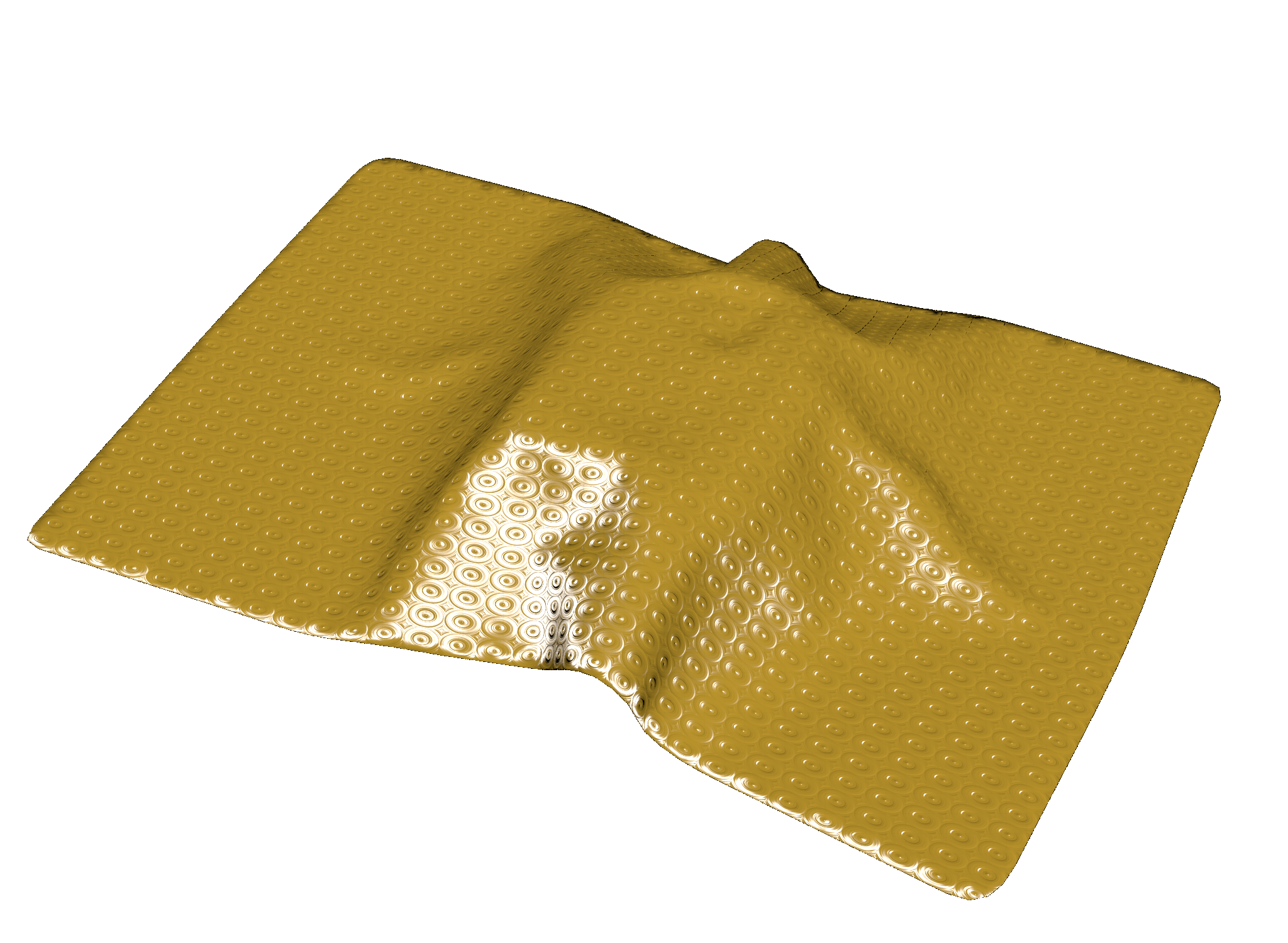
         }&
        \includegraphics[width=0.25\textwidth,trim={0 0 0 7cm},clip]{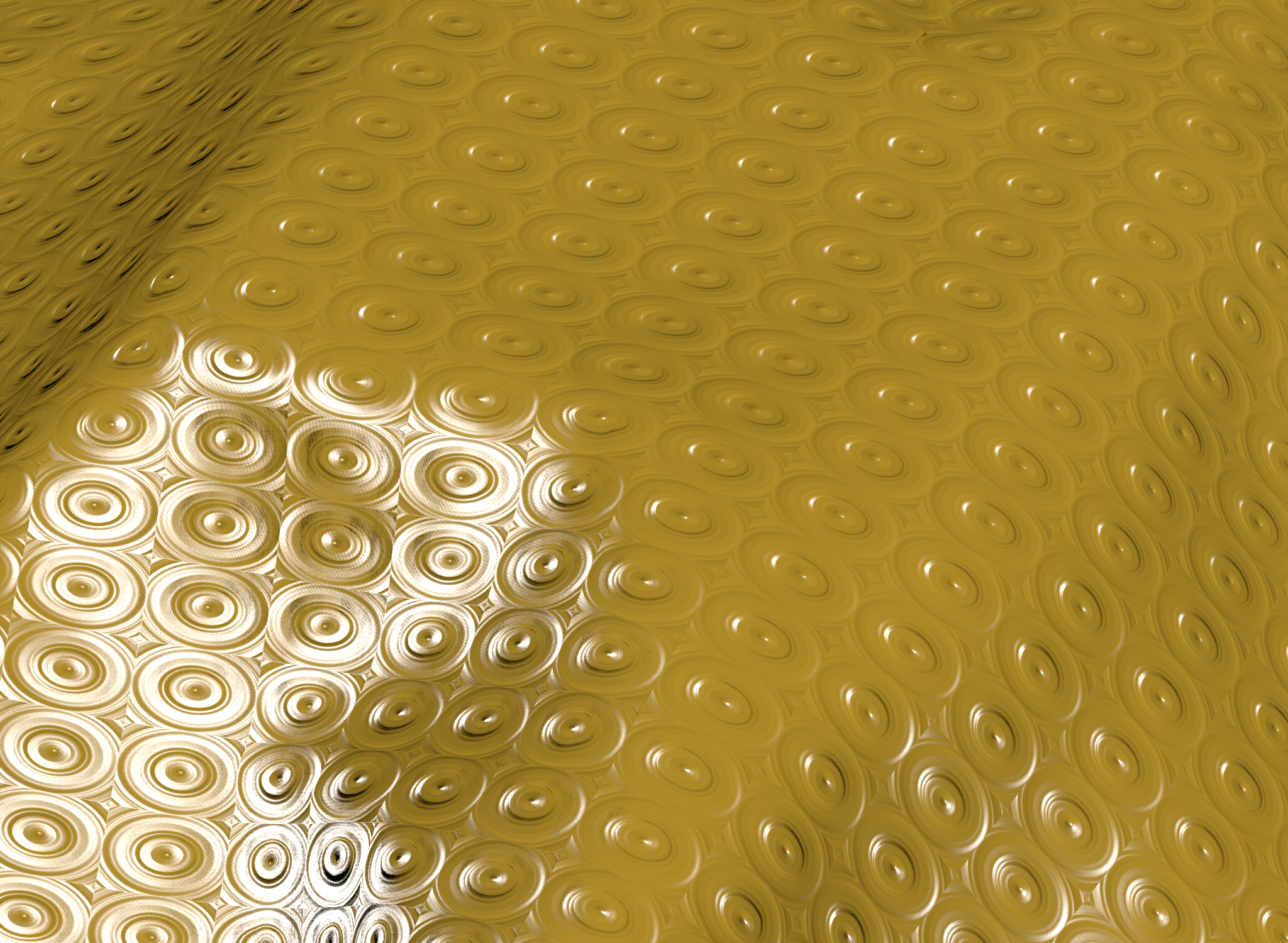}
    \end{tabular}
    \caption{An assortment of materials with our method on a non-flat surface. Please view the video in our supplemental materials for light rotation around the scene as well as gradual zooming in to showcase the level-of-detail. }
    \label{fig_3D}
\end{figure*}
\section{Results}
\label{sec_results}
In what follows, we demonstrate the effectiveness of our method empirically using rendered results.
Specifically, we first show ablation studies to justify the necessity of individual components of our method~(\S\ref{ssec_ablation}).
Then, we compare our model to the NeuMIP baseline using a range of materials~(\S\ref{ssec_evaluation}).

\begin{table*}[t]
\caption{Errors for images rendered across multiple levels of details as shown in \figref{fig_LoD}. }
\centering
\setlength{\tabcolsep}{3pt} 
\begin{tabular}{l|c|c|c|c|c|c|c|c|c|c|c|c|c|c|c|c|c}
& \multicolumn{8}{c|}{Ours MSE} & \multicolumn{8}{c}{NeuMIP MSE} \\\cline{2-9}\cline{10-17}
            {Scene $\downarrow$} \quad {LoD $\rightarrow$}
           & 0 & 1 & 2 & 3 & 4 & 5 & 6 & 7
           & 0 & 1 & 2 & 3 & 4 & 5 & 6 & 7 \\\hline
 Basket cloth  & 0.792& 0.637 & 0.457 & 0.264 & 0.119 & 0.043 & 0.012 & 0.003 &  6.367 & 6.018 & 5.437 & 4.595 & 3.568  &  2.585 & 1.960 & 1.581
\\ \hline
 Metal ring & 0.062&   0.046 & 0.035 &  0.025 & 0.019 & 0.016 & 0.014 & 0.013
& 3.093 &   2.989  & 2.767 & 2.368 & 1.867& 1.305 & 0.802 & 0.448\\ \hline
\end{tabular}
    \label{table_LoD}
\end{table*}

\subsection{Ablation Studies}
\label{ssec_ablation}
In \figref{fig_ablation}, we employ the metal ring and basket cloth examples to demonstrate the significance of each component within our proposed method. The ring example uses displacement maps in which vertical displacement follows a two-dimensional Gaussian function and the basket cloth is rendered with real micro-geometry. As shown in this figure, our hierarchical architecture enabled by the Inception module enhances the overall model's expressive capability. Consistently, our input encoding module and gradient loss prove instrumental in capturing edges and high-frequency features. Furthermore, the remapping strategy aids in the high-quality reconstruction of shadows. In case of skipping the remapping step, the back yarn is missing in the fourth column as marked by the red square. This is due to the using MSE loss where the network tends to evenly reduce the loss in the whole texture. However, the same error has different effects in low-luminance areas and high-luminance areas. The remapping simulates the human eye's reaction to light power, so the network better learns the importance of the texture value. This is a simple example exhibiting strong self-shadowing and sharp highlights.
NeuMIP has difficulties in accurately handling these effects---even when using significantly larger network sizes (see \figref{fig_largeneumip}) that are much more expensive to train and evaluate than our model.

\subsection{Evaluation Results}
\label{ssec_evaluation}

\mypara{Comparisons with large NeuMIP}
In order to challenge the original NeuMIP framework fairly, we also experimented larger size of their network by increasing the number of neurons and layers. As shown in \figref{fig_largeneumip}, the self-shadowing and sharp highlights have difficulty to be captured even with larger size MLP. However, our hierarchical architecture reproduces the features. Additionally, as the network size increases the training and query time are increased while the performance of our method stays nearly the same as the original NeuMIP.

\mypara{Comparisons with previous works}
In  \figref{fig_teaser} and \ref{fig_comparison}, we compare results generated using our method and NeuMIP on a wide range of complex materials (from both our and NeuMIP's datasets). In the Metal ring example, the NeuMIP result misses most shadows and specular highlights. In both the Twill cloth and Victorial fabric samples, NeuMIP has difficulty capturing the sharp highlights correctly. The Metal grid scene showcases high-frequency details that are captured using our hierarchical network architecture while missed by NeuMIP decoder.  We did not compare it to the recent variant of NeuMIP\cite{kuznetsov2022} as they don't publish their code, and they require additional parameters as input (curvature).

The input to our neural method framework is a 7D parameter set that can be obtained using either synthetic datasets or real-measured BTF. We used the leather example from UBO 2014 dataset \cite{UBO2014} to exemplify the effectiveness of our model regardless of the input source. As shown in \figref{fig_comparison}, unlike NeuMIP, our method better captures the fine-grain details as well as the highlights in the leather scene. 

\mypara{Comparisons with large NeuMIP}
In order to challenge the original NeuMIP framework fairly, we also experimented larger size of their network by increasing the number of neurons and layers. As shown in \figref{fig_largeneumip}, the self-shadowing and sharp highlights have difficulty to be captured even with larger size MLP. However, our hierarchical architecture reproduces the features. Additionally, as the network size increases the training and query time are increased while the performance of our method stays nearly the same as the original NeuMIP.

\mypara{Multi-resolution results}
We demonstrate the effectiveness of our method addressing the different levels of detail of the material in \figref{fig_LoD}. As expected, for the courser levels, the errors become smaller as we travel down in the hierarchical structure. This is due to the natural downsampling effect and gradual fading of the high-frequency details. We refer to the closest view as level-0 and the coarser levels are assigned to higher grades. The error scores for different levels are highlighted in \tableref{table_LoD}. In the Basket cloth sample, please note the deeper yarns are missing in NeuMIP while ours can successfully reproduce low-luminance regions as well as the high-frequency features such as edges and fiber details. Furthermore, in \figref{fig_LoD} we show non-flat surface shaded using our method to showcase our integration with a renderer. Please view the accompanying video for the gradual change in the level-of-detail.

\mypara{Quantitative evaluation}
We also measure the numerical error of our neural method when compared to the reference. Our method performance in comparison with NeuMIP is listed in \tableref{table_error} using both MSE loss (Means Square Error) and perceptual loss LPIPS (Learned Perceptual Image Patch Similarity). This shows the overall average on the whole dataset and our method always outperforms NeuMIP using the same configuration. Later, in \tableref{table_LoD} we demonstrate the scores of multiple scales of ours and NeuMIP model using the different levels of detail from the reference dataset.

\section{Discussion and Conclusion}

\mypara{Limitation and future work} 
To integrate our neural reflectance models into physics-based Monte Carlo rendering frameworks, efficient importance sampling techniques for these models need to be developed---which we think is an important problem for future investigation. Our model only captures direct illumination with a single light bounce and the global illumination is another interesting future work. Besides, due to the larger footprint required by the convolution layer. our method is slightly slower than the original method which could be optimized.

Further, adopting our technique to improve the accuracy of the more recent neural reflectance model~\cite{kuznetsov2022} (with better silhouettes) is worth exploring.

Lastly, generalizing our technique to introduce neural BSSRDFs (that can capture subsurface scattering) can be beneficial to many future applications.


\mypara{Conclusion}
In this paper, we improved the accuracy of the NeuMIP~\cite{kuznetsov2021neumip} by introducing a new neural representation as well as a training process for this representation.
Using neural networks with identical sizes, compared with NeuMIP, our neural representation is capable of reproducing detailed specular highlights and shadowing at significantly higher accuracy while better preserving a material's overall color.
Additionally, we proposed an optional modification to the decoder architecture that further enhances the performance.
We demonstrated the effectiveness of our technique by comparing to NeuMIP (at equal network size) using several examples.





\FloatBarrier
\sloppy
\printbibliography                

@article{kuznetsov2021neumip,
author = {Kuznetsov, Alexandr and Mullia, Krishna and Xu, Zexiang and Ha\v{s}an, Milo\v{s} and Ramamoorthi, Ravi},
title = {NeuMIP: multi-resolution neural materials},
year = {2021},
issue_date = {August 2021},
publisher = {Association for Computing Machinery},
address = {New York, NY, USA},
volume = {40},
number = {4},
issn = {0730-0301},
url = {https://doi.org/10.1145/3450626.3459795},
doi = {10.1145/3450626.3459795},
journal = {ACM Trans. Graph.},
month = {jul},
articleno = {175},
numpages = {13}
}

@inproceedings{kuznetsov2022,
author = {Kuznetsov, Alexandr and Wang, Xuezheng and Mullia, Krishna and Luan, Fujun and Xu, Zexiang and Hasan, Milos and Ramamoorthi, Ravi},
title = {Rendering Neural Materials on Curved Surfaces},
year = {2022},
isbn = {9781450393379},
publisher = {Association for Computing Machinery},
address = {New York, NY, USA},
url = {https://doi.org/10.1145/3528233.3530721},
doi = {10.1145/3528233.3530721},
booktitle = {ACM SIGGRAPH 2022 Conference Proceedings},
articleno = {9},
numpages = {9},
location = {Vancouver, BC, Canada},
series = {SIGGRAPH '22}
}

@article{khungurn2015matching,
author = {Khungurn, Pramook and Schroeder, Daniel and Zhao, Shuang and Bala, Kavita and Marschner, Steve},
title = {Matching Real Fabrics with Micro-Appearance Models},
year = {2016},
issue_date = {December 2015},
publisher = {Association for Computing Machinery},
address = {New York, NY, USA},
volume = {35},
number = {1},
issn = {0730-0301},
url = {https://doi.org/10.1145/2818648},
doi = {10.1145/2818648},
journal = {ACM Trans. Graph.},
month = {dec},
articleno = {1},
numpages = {26}
}

@article{Montazeri2021mechanics,
author = {Montazeri, Zahra and Xiao, Chang and Fei, Yun and Zheng, Changxi and Zhao, Shuang},
title = {Mechanics-Aware Modeling of Cloth Appearance},
year = {2021},
issue_date = {Jan. 2021},
publisher = {IEEE Educational Activities Department},
address = {USA},
volume = {27},
number = {1},
issn = {1077-2626},
url = {https://doi.org/10.1109/TVCG.2019.2937301},
doi = {10.1109/TVCG.2019.2937301},
journal = {IEEE Transactions on Visualization and Computer Graphics},
month = {jan},
pages = {137–150},
numpages = {14}
}

@article{montazeri2021practical,
author = {Montazeri, Zahra and Gammelmark, S\o{}ren B. and Zhao, Shuang and Jensen, Henrik Wann},
title = {A practical ply-based appearance model of woven fabrics},
year = {2020},
issue_date = {December 2020},
publisher = {Association for Computing Machinery},
address = {New York, NY, USA},
volume = {39},
number = {6},
issn = {0730-0301},
url = {https://doi.org/10.1145/3414685.3417777},
doi = {10.1145/3414685.3417777},
journal = {ACM Trans. Graph.},
month = {nov},
articleno = {251},
numpages = {13}
}

@article{montazeri2020practical,
author = {Montazeri, Zahra and Gammelmark, S\o{}ren B. and Zhao, Shuang and Jensen, Henrik Wann},
title = {A practical ply-based appearance model of woven fabrics},
year = {2020},
issue_date = {December 2020},
publisher = {Association for Computing Machinery},
address = {New York, NY, USA},
volume = {39},
number = {6},
issn = {0730-0301},
url = {https://doi.org/10.1145/3414685.3417777},
doi = {10.1145/3414685.3417777},
journal = {ACM Trans. Graph.},
month = {nov},
articleno = {251},
numpages = {13}
}

@INPROCEEDINGS {inception,
author = {C. Szegedy and Wei Liu and Yangqing Jia and P. Sermanet and S. Reed and D. Anguelov and D. Erhan and V. Vanhoucke and A. Rabinovich},
booktitle = {2015 IEEE Conference on Computer Vision and Pattern Recognition (CVPR)},
title = {Going deeper with convolutions},
year = {2015},
volume = {},
issn = {1063-6919},
pages = {1-9},
doi = {10.1109/CVPR.2015.7298594},
url = {https://doi.ieeecomputersociety.org/10.1109/CVPR.2015.7298594},
publisher = {IEEE Computer Society},
address = {Los Alamitos, CA, USA},
month = {jun}
}

@article{mildenhall2020nerf,
  title={Nerf: Representing scenes as neural radiance fields for view synthesis},
  author={Mildenhall, Ben and Srinivasan, Pratul P and Tancik, Matthew and Barron, Jonathan T and Ramamoorthi, Ravi and Ng, Ren},
  journal={Communications of the ACM},
  volume={65},
  number={1},
  pages={99--106},
  year={2021},
  publisher={ACM New York, NY, USA}
}

@article{thonat2021,
author = {Thonat, Theo and Beaune, Francois and Sun, Xin and Carr, Nathan and Boubekeur, Tamy},
title = {Tessellation-Free Displacement Mapping for Ray Tracing},
year = {2021},
issue_date = {December 2021},
publisher = {Association for Computing Machinery},
address = {New York, NY, USA},
volume = {40},
number = {6},
issn = {0730-0301},
url = {https://doi.org/10.1145/3478513.3480535},
doi = {10.1145/3478513.3480535},
journal = {ACM Trans. Graph.},
month = {dec},
articleno = {282},
numpages = {16}
}

@article{dana1999,
author = {Dana, Kristin J. and van Ginneken, Bram and Nayar, Shree K. and Koenderink, Jan J.},
title = {Reflectance and Texture of Real-World Surfaces},
year = {1999},
issue_date = {Jan. 1999},
publisher = {Association for Computing Machinery},
address = {New York, NY, USA},
volume = {18},
number = {1},
issn = {0730-0301},
url = {https://doi.org/10.1145/300776.300778},
doi = {10.1145/300776.300778},
journal = {ACM Trans. Graph.},
month = {jan},
pages = {1–34},
numpages = {34}
}

@inproceedings{oliveira2000relief,
author = {Oliveira, Manuel M. and Bishop, Gary and McAllister, David},
title = {Relief texture mapping},
year = {2000},
isbn = {1581132085},
publisher = {ACM Press/Addison-Wesley Publishing Co.},
address = {USA},
url = {https://doi.org/10.1145/344779.344947},
doi = {10.1145/344779.344947},
booktitle = {Proceedings of the 27th Annual Conference on Computer Graphics and Interactive Techniques},
pages = {359–368},
numpages = {10},
series = {SIGGRAPH '00}
}

@inproceedings{kaneko2001parallax,
  title={Detailed shape representation with parallax mapping},
  author={Kaneko, Tomomichi and Takahei, Toshiyuki and Inami, Masahiko and Kawakami, Naoki and Yanagida, Yasuyuki and Maeda, Taro and Tachi, Susumu},
  booktitle={Proceedings of ICAT},
  volume={2001},
  pages={205--208},
  year={2001}
}

@article{policarpo2005relief,
author = {Policarpo, F\'{a}bio and Oliveira, Manuel M. and Comba, Jo\~{a}o L. D.},
title = {Real-time relief mapping on arbitrary polygonal surfaces},
year = {2005},
issue_date = {July 2005},
publisher = {Association for Computing Machinery},
address = {New York, NY, USA},
volume = {24},
number = {3},
issn = {0730-0301},
url = {https://doi.org/10.1145/1073204.1073292},
doi = {10.1145/1073204.1073292},
journal = {ACM Trans. Graph.},
month = {jul},
pages = {935},
numpages = {1}
}

@article{wang2003vdm,
  title={View-dependent displacement mapping},
  author={Wang, Lifeng and Wang, Xi and Tong, Xin and Lin, Stephen and Hu, Shimin and Guo, Baining and Shum, Heung-Yeung},
  journal={ACM Transactions on graphics (TOG)},
  volume={22},
  number={3},
  pages={334--339},
  year={2003},
  publisher={ACM New York, NY, USA}
}

@inproceedings {wang2004gdm,
booktitle = {Eurographics Workshop on Rendering},
editor = {Alexander Keller and Henrik Wann Jensen},
title = {{Generalized Displacement Maps}},
author = {Wang, Xi and Tong, Xin and Lin, Stephen and Hu, Shimin and Guo, Baining and Shum, Heung-Yeung},
year = {2004},
publisher = {The Eurographics Association},
ISSN = {1727-3463},
ISBN = {3-905673-12-6},
DOI = {10.2312/EGWR/EGSR04/227-233}
}

@article {filip2008compression,
journal = {Computer Graphics Forum},
title = {{Bidirectional Texture Function Compression Based on Multi-Level Vector Quantization}},
author = {Havran, V. and Filip, J. and Myszkowski, K.},
year = {2010},
publisher = {The Eurographics Association and Blackwell Publishing Ltd},
ISSN = {1467-8659},
DOI = {10.1111/j.1467-8659.2009.01585.x}
}

@article{wang2005mdf,
author = {Yu, Yizhou and Zhou, Kun and Xu, Dong and Shi, Xiaohan and Bao, Hujun and Guo, Baining and Shum, Heung-Yeung},
title = {Mesh editing with poisson-based gradient field manipulation},
year = {2004},
issue_date = {August 2004},
publisher = {Association for Computing Machinery},
address = {New York, NY, USA},
volume = {23},
number = {3},
issn = {0730-0301},
url = {https://doi.org/10.1145/1015706.1015774},
doi = {10.1145/1015706.1015774},
journal = {ACM Trans. Graph.},
month = {aug},
pages = {644–651},
numpages = {8}
}

@inproceedings{rainer2019autoencoder,
  title={Neural BTF compression and interpolation},
  author={Rainer, Gilles and Jakob, Wenzel and Ghosh, Abhijeet and Weyrich, Tim},
  booktitle={Computer Graphics Forum},
  volume={38},
  number={2},
  pages={235--244},
  year={2019},
  organization={Wiley Online Library}
}

@INPROCEEDINGS{zhang2021nerfw,
  author={Zhang, Jian and Zhang, Yuanqing and Fu, Huan and Zhou, Xiaowei and Cai, Bowen and Huang, Jinchi and Jia, Rongfei and Zhao, Binqiang and Tang, Xing},
  booktitle={2022 IEEE/CVF Conference on Computer Vision and Pattern Recognition (CVPR)}, 
  title={Ray Priors through Reprojection: Improving Neural Radiance Fields for Novel View Extrapolation}, 
  year={2022},
  volume={},
  number={},
  pages={18355-18365},
  doi={10.1109/CVPR52688.2022.01783}}

@INPROCEEDINGS{wang2022fourier,
  author={Wang, Liao and Zhang, Jiakai and Liu, Xinhang and Zhao, Fuqiang and Zhang, Yanshun and Zhang, Yingliang and Wu, Minve and Yu, Jingyi and Xu, Lan},
  booktitle={2022 IEEE/CVF Conference on Computer Vision and Pattern Recognition (CVPR)}, 
  title={Fourier PlenOctrees for Dynamic Radiance Field Rendering in Real-time}, 
  year={2022},
  volume={},
  number={},
  pages={13514-13524},
  doi={10.1109/CVPR52688.2022.01316}}

@article{sobel,
  title={A 3x3 Isotropic Gradient Operator for Image Processing},
  author={I. Sobel and G. Feldman},
  journal={Pattern Classification and Scene Analysis},
  pages={pp. 271-272},
  year={1973},
}

@article{canny,
  title={A computational approach to edge detection},
  author={Canny, John},
  journal={IEEE Transactions on pattern analysis and machine intelligence},
  number={6},
  pages={679--698},
  year={1986},
  publisher={Ieee}
}

@article{Mitsuba,
  title={Mitsuba 2: A retargetable forward and inverse renderer},
  author={Nimier-David, Merlin and Vicini, Delio and Zeltner, Tizian and Jakob, Wenzel},
  journal={ACM Transactions on Graphics (TOG)},
  volume={38},
  number={6},
  pages={1--17},
  year={2019},
  publisher={ACM New York, NY, USA}
}

@misc{keyshot,
  title={KeyShot},
  author={Luxion},
  howpublished={\url{https://www.keyshot.com}},
  year={2020}
}

@article{UBO2014,
  title={Material Classification based on Training Data Synthesized Using a BTF Database},
  author={Michael Weinmann, Juergen Gall and Reinhard Klein.},
  journal={ECCV},
  year={2014},
  publisher={ECCV}
}

@book{fourier,
  title={The Fourier transform and its applications},
  author={Bracewell, Ronald Newbold and Bracewell, Ronald N},
  volume={31999},
  year={1986},
  publisher={McGraw-Hill New York}
}

@misc{
rahaman2019,
title={On the Spectral Bias of Neural Networks},
author={Nasim Rahaman and Aristide Baratin and Devansh Arpit and Felix Draxler and Min Lin and Fred Hamprecht and Yoshua Bengio and Aaron Courville},
year={2019},
url={https://openreview.net/forum?id=r1gR2sC9FX},
}

@article{bertalmio2020evidence,
  title={Evidence for the intrinsically nonlinear nature of receptive fields in vision},
  author={Bertalmío, M. and Martín, A. and Kane, D. and Malo, J.},
  journal={Scientific Reports},
  volume={10},
  number={1},
  pages={1--15},
  year={2020},
  url={https://doi.org/10.1038/s41598-020-73113-0}
}

@article{zeltner2023real,
author = {Zeltner, Tizian and Rousselle, Fabrice and Weidlich, Andrea and Clarberg, Petrik and Nov\'{a}k, Jan and Bitterli, Benedikt and Evans, Alex and Davidovi\v{c}, Tom\'{a}\v{s} and Kallweit, Simon and Lefohn, Aaron},
title = {Real-Time Neural Appearance Models},
year = {2024},
publisher = {Association for Computing Machinery},
address = {New York, NY, USA},
issn = {0730-0301},
url = {https://doi.org/10.1145/3659577},
doi = {10.1145/3659577},
note = {Just Accepted},
journal = {ACM Trans. Graph.},
month = {apr}
}

@inproceedings{fan2022neural,
author = {Fan, Jiahui and Wang, Beibei and Hasan, Milos and Yang, Jian and Yan, Ling-Qi},
title = {Neural Layered BRDFs},
year = {2022},
isbn = {9781450393379},
publisher = {Association for Computing Machinery},
address = {New York, NY, USA},
url = {https://doi.org/10.1145/3528233.3530732},
doi = {10.1145/3528233.3530732},
booktitle = {ACM SIGGRAPH 2022 Conference Proceedings},
articleno = {4},
numpages = {8},
location = {Vancouver, BC, Canada},
series = {SIGGRAPH '22}
}

@inproceedings{rainer2020unified,
  title={Unified neural encoding of BTFs},
  author={Rainer, Gilles and Ghosh, Abhijeet and Jakob, Wenzel and Weyrich, Tim},
  booktitle={Computer Graphics Forum},
  volume={39},
  number={2},
  pages={167--178},
  year={2020},
  organization={Wiley Online Library}
}

@inproceedings{xu2023neusample,
author = {Xu, Bing and Wu, Liwen and Hasan, Milos and Luan, Fujun and Georgiev, Iliyan and Xu, Zexiang and Ramamoorthi, Ravi},
title = {NeuSample: Importance Sampling for Neural Materials},
year = {2023},
isbn = {9798400701597},
publisher = {Association for Computing Machinery},
address = {New York, NY, USA},
url = {https://doi.org/10.1145/3588432.3591524},
doi = {10.1145/3588432.3591524},
booktitle = {ACM SIGGRAPH 2023 Conference Proceedings},
articleno = {41},
numpages = {10},
location = {<conf-loc>, <city>Los Angeles</city>, <state>CA</state>, <country>USA</country>, </conf-loc>},
series = {SIGGRAPH '23}
}

@article{rodriguez2023neubtf,
  title={NeuBTF: Neural fields for BTF encoding and transfer},
  author={Rodriguez-Pardo, Carlos and Kazatzis, Konstantinos and Lopez-Moreno, Jorge and Garces, Elena},
  journal={Computers \& Graphics},
  year={2023},
  publisher={Elsevier}
}

@article{gauthier2022mipnet,
author = {Gauthier, Alban and Faury, Robin and Levallois, J\'{e}r\'{e}my and Thonat, Th\'{e}o and Thiery, Jean-Marc and Boubekeur, Tamy},
title = {MIPNet: Neural Normal-to-Anisotropic-Roughness MIP Mapping},
year = {2022},
issue_date = {December 2022},
publisher = {Association for Computing Machinery},
address = {New York, NY, USA},
volume = {41},
number = {6},
issn = {0730-0301},
url = {https://doi.org/10.1145/3550454.3555487},
doi = {10.1145/3550454.3555487},
journal = {ACM Trans. Graph.},
month = {nov},
articleno = {246},
numpages = {12}
}

@inproceedings{baatz2022nerf,
  title={NeRF-Tex: Neural Reflectance Field Textures},
  author={Baatz, Hendrik and Granskog, Jonathan and Papas, Marios and Rousselle, Fabrice and Nov{\'a}k, Jan},
  booktitle={Computer Graphics Forum},
  volume={41},
  number={6},
  pages={287--301},
  year={2022},
  organization={Wiley Online Library}
}

@INPROCEEDINGS {xiang2021neutex,
author = {F. Xiang and Z. Xu and M. Hasan and Y. Hold-Geoffroy and K. Sunkavalli and H. Su},
booktitle = {2021 IEEE/CVF Conference on Computer Vision and Pattern Recognition (CVPR)},
title = {NeuTex: Neural Texture Mapping for Volumetric Neural Rendering},
year = {2021},
volume = {},
issn = {},
pages = {7115-7124},
doi = {10.1109/CVPR46437.2021.00704},
url = {https://doi.ieeecomputersociety.org/10.1109/CVPR46437.2021.00704},
publisher = {IEEE Computer Society},
address = {Los Alamitos, CA, USA},
month = {jun}
}

\end{document}